\documentclass[aps, pre, reprint, superscriptaddress, longbibliography, showkeys, nofootinbib, floatfix]{revtex4-1}

\usepackage{amssymb, mathtools, amsmath, amsfonts, physics, blkarray, booktabs}
\usepackage{graphicx, color, soul, cancel, lipsum, xspace, tikz, placeins}
\usepackage[colorlinks=true,linkcolor=blue,urlcolor=blue,citecolor=blue,pdfusetitle]{hyperref}

\newcommand{\transR}{+}
\newcommand{\transL}{-}
\newcommand{\traff}{\langle K\rangle}

\newcommand{\lbra}[1]{\langle \hspace{-.7ex} \langle #1 \vert \hspace{-.45ex} \vert}
\newcommand{\lket}[1]{ \vert \hspace{-.45ex} \vert #1 \rangle \hspace{-.7ex} \rangle}
\newcommand{\lbraket}[2]{\langle \hspace{-.7ex} \langle #1 \vert \hspace{-.45ex} \vert #2 \rangle \hspace{-.7ex} \rangle}

\definecolor{red}{rgb}{0,0,0}

\begin{document}

\title{What to learn from {\color{red}a} few visible transitions' statistics?}
\author{Pedro E. Harunari}
\email{pedroharunari@gmail.com}
\affiliation{Instituto de F\'isica da Universidade de S\~ao Paulo, 05314-970 S\~ao Paulo, Brazil}
\affiliation{Complex Systems and Statistical Mechanics, Department of Physics and Materials Science,
University of Luxembourg, L-1511 Luxembourg, Luxembourg}

\author{Annwesha Dutta}
\affiliation{ICTP -- The Abdus Salam International Centre for Theoretical Physics, Strada Costiera 11, 34151 Trieste, Italy}
\affiliation{Department of Physics,  Indian Institute of Science Education and Research, Tirupati 517507, India}

\author{Matteo Polettini}
\affiliation{Complex Systems and Statistical Mechanics, Department of Physics and Materials Science,
University of Luxembourg, L-1511 Luxembourg, Luxembourg}

\author{\'Edgar Rold\'an}
\email{edgar@ictp.it}
\affiliation{ICTP -- The Abdus Salam International Centre for Theoretical Physics, Strada Costiera 11, 34151 Trieste, Italy}

\date{\today} 
\begin{abstract}
Interpreting partial information collected from systems subject to noise is a key problem across scientific disciplines. Theoretical frameworks often focus on the dynamics of variables that result from coarse-graining the internal states of a physical system. However, most experimental apparatuses  {\color{red}can only detect a partial set of  transitions}, while internal states of the physical system are blurred or inaccessible. Here, we consider an observer who records a time series of occurrences of one or several transitions performed by a system, under the assumption that its underlying dynamics is Markovian. We pose the question of how one can use the transitions' information to make inferences of dynamical, thermodynamical, and biochemical properties. First, {\color{red}elaborating on} first-passage time techniques, we derive analytical expressions for the probabilities of consecutive transitions and for the time elapsed between them, which we call {\em inter-transition times}. Second, we {\color{red}derive a lower bound for the   entropy production rate  that equals to the sum of two}  non-negative contributions, one due to the statistics of transitions and a second due to the statistics of inter-transition times. We also show that when only one current is measured, our estimate still detects irreversibility even in the absence of net currents in the transition time series. Third, we verify our results with numerical simulations {\color{red}using  unbiased estimates of entropy production, which we make available as an open-source toolbox. We illustrate  the developed framework in experimentally-validated  biophysical models of kinesin and dynein molecular motors, and in a minimal model for} template-directed polymerization. Our numerical results reveal that while entropy production is entailed in the statistics of two successive  transitions of the same type (i.e. repeated transitions),  the statistics of two different successive transitions (i.e. alternated transitions) can probe the existence  of an underlying disorder in the motion of a molecular motor. {\color{red} Taken all together, our results highlight the power of inference from transition statistics ranging from thermodynamic quantities to network-topology properties of Markov processes.}
\end{abstract}
\maketitle

%
\section{Introduction}
%
Model systems in physics \cite{vankampen1992}, chemistry \cite{tamir1998,anderson2011, Avanzini2021},  biology \cite{allen2010,Kolomeisky2007,Chowdhury2013}, and computation~\cite{wolpert2019stochastic} are routinely described by Markov processes, which are also amenable to thermodynamic analysis \cite{PhysRevE.81.051133, PhysRevE.82.011143, PhysRevE.82.011144, 10.1143/PTPS.130.17, PhysRevE.82.021120}. This approach thrives when there is full knowledge of the system's internal state, but in most practical applications experimental apparatuses access few degrees of freedom or have a finite resolution, thus only partial information is available. One example is the rotation of flagella in a bacterial motor \cite{PhysRevLett.96.058105}: observation of orientation switches in the direction of the bacteria's flagella suggests the existence of internal states that are hidden from the observer.

\begin{figure}[ht!]
    \centering
    \includegraphics[width=1\columnwidth]{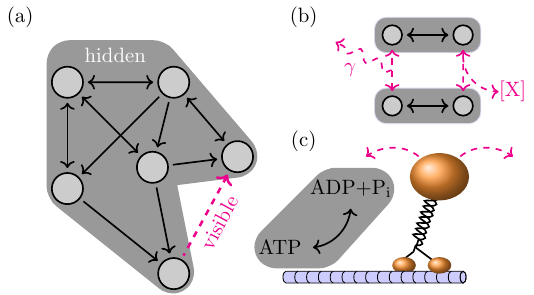}
    \caption{Illustrations of the partial information acquired by an apparatus that can only detect a few visible transitions. (a) Setup of our framework, for the case of only one visible transition. A physical system performs a Markov-jump process in a network of states (circles)  all of which are hidden from an external observer. The observer can only see occurrences of {\color{red}a} few visible transitions (one in this example), while the rest of the transitions remain hidden during the data acquisition (black shaded area). (b,c) Model examples described by our theory: (b) photon emission/absorption \(\gamma\) or synthesis/consumption of chemical species $X$ that signal the occurrence of some transitions, and (c) {\color{red} a molecular motor performing steps along a track and  transitions related to spatial motion along the track are detected by monitoring the position of the  cargo (orange sphere) while chemical fuel consumption (ATP hydrolysis) remains often hidden (grey box)}, see also Fig.~\ref{fig:summary_pic}.}
    \label{fig:hidden}
\end{figure}

The problem of measuring partial information, or of coarse-graining degrees of freedom, is usually framed in terms of the internal state of a system \cite{bo2017multiple,esposito2012stochastic, pigolotti_vulpiani_2008, Rahav_2007, PhysRevLett.125.110601, PhysRevX.8.031038}. However, in most practical applications, an external observer only measures ``footprints'' of one or several transitions, rather than the internal state itself, as sketched in Fig.~\ref{fig:hidden}(a). These footprints may be due to physical degrees of freedom satisfying microscopic reversibility, in which case it is possible to talk about their energetic and entropic balance, as sketched in Fig.~\ref{fig:hidden}(b) where the observer can detect the emission and absorption of a photon $\gamma$, or the production or consumption of a chemical species $X$. Finally, Fig.~\ref{fig:hidden}(c) sketches the motion of a molecular motor (e.g. a kinesin) along a periodic track (e.g. microtubule). The motor undergoes structural changes followed by a translocation step associated to the consumption of some resources (e.g. adenosine triphosphate). The only visible transitions are in this case the forward and backward steps along the track. {\color{red} As explained below, this situation is customary in experiments where the motion of a microscopic bead attached to the motor can be used to detect spatial displacements along the track while conformational changes and chemical fuel consumption remain undetectable to the experimenter~\cite{vale1985identification}. }

{\color{red} Significant developments in single-molecule experimental techniques with biological systems at cellular and sub-cellular level have been reported over the last  few decades~\cite{Zlatanova2006}. For example, the motion  of biomolecular machines involved in cellular transport such as kinesin \cite{Verbrugge2007}, dynein \cite{Ananthanarayanan2015,Niekamp2021} and myosin \cite{Desai2015} has been resolved at the sub-nanometer resolution. Examples include real-time tracking  of individual, fluorescently-tagged biomolecules~\cite{Moerner2003,Joo2008} followed by data analysis techniques of the recorded trajectories using e.g. kymographs \cite{Mangeol2016,ReckPeterson2006}. In most of these experiments, biomolecular machines are subject to nonequilibrium forces that may be intrinsic (e.g.  chemical reactions) or extrinsic (e.g. mechanical forces exerted by optical tweezers). 
This motivates the fact that the motion of the molecular motor is routinely described by Markovian nonequilibrium stationary states. 



The typical scenario of  single-molecule studies is such that only a partial set of degrees of freedom and/or transitions  are experimentally accessible. For example, using high-resolution optical tweezers it is customary that the spatial transitions (e.g. a step in a linear track) can be measured experimentally while conformational changes or chemical reactions remain hidden to the experimenter. 
This is the case of e.g. the molecular machines of the central dogma of genetic information processing, DNA polymerase \cite{Zlatanova2006},  RNA polymerase \cite{Abbondanzieri2005}, and ribosomes \cite{Aitken2010,Wen2008}.
Because every transition during molecular motor motion is accompanied by changes in internal energy due to the chemical energy arising from the coupling of the system to chemical reservoirs~\cite{Dutta2020,Lipowsky2009}, having reliable estimates of entropy production from the observation of a partial set of transitions is key to develop accurate  bounds on efficiency and thermodynamic costs of molecular machines~\cite{Skinner2021,Otsubo2022}.


In an attempt to extract useful thermodynamic information  from  the partial observation of {\color{red}a} few visible transitions' statistics, we develop a transition-based coarse-graining framework for  continuous-time Markov processes. Our analytical progress leads to descriptions and predictions suitable to systems whose available information comes only from counting transitions, and measuring the time elapsed between two consecutive transitions ---a key concept that we denote as {\em inter-transition times}. In particular we focus on how can one infer  thermodynamic and topological properties from the sole observation of 
 inter-transition times and frequencies of transitions, and what are the  consequences for experimentally-validated  models of biomolecular systems?

\section{goals and main results}
}

Recent work revealed that information extracted from transitions between {\color{red}a} few selected visible states provides information about entropy production~\cite{polettini2019effective, martinez2019inferring, PhysRevE.91.012130}. Yet, most of these efforts rely on knowledge about the internal states of the system. Instead, the main question we address in this contribution is: what can be learnt about a system solely from the occurrence of {\color{red}a} few visible transitions (denoted $\ell_i \in \mathcal{L}$) and from the time elapsed between them (denoted $t_i$ {\color{red}and called {\em inter-transition time}})? Our object of study is therefore a time series of the form
\FloatBarrier
\begin{equation}\label{trajectory}
    \includegraphics[width=.8\columnwidth]{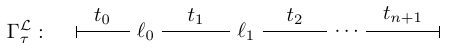}
\end{equation}
\FloatBarrier
\noindent
where $t_i$ denotes the time elapsed between the occurrence of two successive transitions $\ell_{i-1}, \ell_{i}\in \mathcal{L}$, with~$\ell_0$ the  first transition observed. {\color{red} Notice that the subindex  $\tau=\sum_{i=0}^{n+1} t_i $ in $\Gamma^\mathcal{L}_\tau$ indicates the total time duration of the observed trajectory which is a deterministic quantity, whereas all $t_i$ are all positive random variables. }

In the following, we use Dirac's notation for vectors where \(\ket{i}\) is a column vector with entries \(\delta_{i,j}\) for \(j\) spanning through the state space, so that for example $\bra{i} \mathbf{A} \ket{j}=A_{ij}$ is the $i$-th row and $j$-th column entry of matrix $\mathbf{A}$. We introduce a special notation when we deal with transitions \(\ell \in \mathcal{L}\): 
$\lbra{\ell} $ is a row vector that has all zero entries except the element corresponding to source state of transition $\ell$. On the other hand, $\lket{\ell}$ is a column vector that has all zero entries except  the element corresponding to the target state of transition $\ell$.  For example  transition $\ell=1\to 3$ has $\lbra{\ell}=\bra{1}$ and $\lket{\ell}=\ket{3}$, and the matrix element associated with transition $1\to 3$  is \( \lbra{\ell} \textbf{A}^\mathsf{T} \lket{\ell} =\bra{3} \mathbf{A}\ket{1}={A}_{3,1}\), where~$\mathsf{T}$ denotes matrix transposition\footnote{Notice that for any two states $i$ and $j$ we have $\langle i | j \rangle = \delta_{i,j}$, with $\delta_{i,j}$ Kronecker's delta, whereas for transitions~$\lbraket{\ell}{\ell} \neq 1$.}.

We assume that the underlying (hidden) dynamics that produces the collected data is a continuous-time, discrete-state space Markov process with time independent rates (also known as jump process) over an irreducible network (from now on simply called Markov chain). The time series $\Gamma_\tau^\mathcal{L}$ is reminiscent of so-called hidden Markov processes, but we emphasize again that here focus is on visible transitions rather than visible states. We focus on the following statistical quantities, which are easily accessible in experimental settings:

\begin{itemize}

    \item Histograms collecting the frequency $F(t | \ell_i, \ell_{i+1}) dt$ that the time $t_i$ elapsed between  $\ell_{i}$ and $\ell_{i+1}$, called {\em inter-transition time}, lies within the interval $[t, t+dt)$.

    \item The conditional frequency $F(\ell_{i+1}| \ell_i)$ that a transition $\ell_{i+1}$ is observed, given that the previous was $\ell_i$. We call the case $\ell_i=\ell_{i+1}$ as {\em repeated} transitions and the case $\ell_i \neq\ell_{i+1}$ as {\em alternated} transitions.

    \item The frequency that a transition $\ell$ occurs in an observed trajectory $F(\ell)$.

\end{itemize}

In this paper we characterize these quantities from a statistical, a thermodynamic, and a biophysical point of view. The first task (statistical) is important from a fundamental point of view, to understand which features from a hidden process can be learnt by looking only at the statistics of {\color{red}a} few visible transitions. We focus the second (thermodynamic) task on inferring the rate of entropy production of the underlying Markov chain, which is a key quantity to characterize the irreversibility of a nonequilibrium process. The third (biophysical) task is important from an applied point of view,  because  most single-molecule experiments retrieve  partial information about the nonequilibrium dynamics of biological systems. 

To tackle these objectives, we derive analytical expressions for the expected value of the three aforementioned transition statistics. From the thermodynamic point of view, on the additional assumption that for every visible transition $\ell \in \mathcal{L}$ its reversed $\overline{\ell} \in \mathcal{L}$ is also visible -- which we dub {\it visible reversibility} -- we compute and characterize the visible  stationary rate of entropy production 
\begin{equation}\label{eq:sigmaL}
\sigma_{\mathcal{L}}=\lim_{\tau\to\infty} \frac{1}{\tau} D\left( P[ \Gamma_\tau^\mathcal{L} ] ||P[ \overline{\Gamma}_\tau^\mathcal{L} ] \right),
\end{equation}
defined as the rate of Kullback-Leibler divergence\footnote{We denote by  $D [ P(x) \vert\vert Q(x) ] = \int_0^\infty \mathrm{d}x P(x)\ln [P(x)/Q(x)]\geq 0$ the Kullback-Leibler divergence between the probability distributions $P$ and $Q$ of the random variable $x$~\cite{Cover2006}. This information-theoretic measure can be generalized to distributions of multiple random variables and path probabilities of stochastic processes, see e.g.~\cite{PhysRevE.85.031129, PhysRevLett.98.080602, Parrondo_2009} for applications in stochastic thermodynamics.}
\begin{equation}D(P[\Gamma_\tau^\mathcal{L}] ||P[\overline{\Gamma}_\tau^\mathcal{L}])=\int \mathcal{D}\Gamma_\tau^\mathcal{L} P[\Gamma_\tau^\mathcal{L}]\ln (P[\Gamma_\tau^\mathcal{L}]/P[\overline{\Gamma}_\tau^\mathcal{L}])\end{equation}
of the probability density of $\Gamma_\tau^\mathcal{L}$ with respect to that of its suitably-defined time-reversed trajectory $\overline{\Gamma}_\tau^\mathcal{L}$.
Finally, we apply the formalism to stochastic models of the molecular motor motion of dynein, kinesin, and polymerization in disordered tracks. 

Our main results are:
\begin{itemize}
    \item[I)] {\color{red} {\it Analytical expressions for inter-transition time probabilities in terms of parameters of the hidden Markov chain}. To this aim, we solve analytically a first-passage time problem in transition space, i.e. a ``first-transition time'' problem \cite{firsttransitiontimes}. More specifically}, letting $\mathbf{W}$ be any transition rate matrix (generator of a Markov chain), we introduce a {\em survival matrix} $\mathbf{S}$ obtained  by setting the entries in~$\mathbf{W}$ corresponding to the visible transitions to zero (see Eq.~\eqref{eq:Sdef} for a rigorous definition). Mapping the occurrence of transitions to a first-passage-time problem, for the probability density of transition $\ell_{i+1}$ happening in the infinitesimal time interval $[t, t+dt)$, {\color{red}and} given that the previous visible transition was $\ell_i${\color{red}, we find}
    \begin{equation}\label{resultintro}
        P(t, \ell_{i+1}\vert \ell_i) = - \lbra{\ell_{i+1}} \mathbf{W}^{\mathsf{T}} \lket{\ell_{i+1}} \lbra{\ell_{i+1}} \exp(t\mathbf{S}) \lket{\ell_{i}}.
    \end{equation}
    The first factor \(\lbra{\ell_{i+1}} \mathbf{W}^{\mathsf{T}} \lket{\ell_{i+1}}\) is the rate of transition \(\ell_{i+1}\), and the second factor \(\lbra{\ell_{i+1}} \exp(t\mathbf{S}) \lket{\ell_{i}}\) is the probability of going from state \(\lket{\ell_{i}}\) to \(\lbra{\ell_{i+1}}\) in time \(t\) without performing any visible transition.
    
    From {\color{red}Eq.}~\eqref{resultintro} we obtain an explicit expression for the conditional probability of successive transitions, the inter-transition time probability density and the probability of the next observed transition given the current occupation distribution. Furthermore we provide explicit expressions in the case of hidden state spaces with ring topology, which we validate with {\color{red}the above} analytical expression~\eqref{resultintro}.
    Eq.~\eqref{resultintro} generalizes results in first-passage time problems from reaching a subset of states ~\cite{redner2001guide,sekimoto2021derivation} to  performing an arbitrary subset of transitions.

    \item[II)] {\it Assuming visible reversibility}{\color{red}, the visibility of the opposite of each visible transition, we calculate the stationary rate of entropy production $\sigma_\mathcal{L}$ given by Eq.~\eqref{eq:sigmaL} and compare it with that of the hidden Markov chain, $\sigma$.} In particular we prove that
    \begin{align}
    \sigma_{\mathcal{L}} \leq \sigma ,
    \label{eq:5}
    \end{align}
    with the equality holding for systems with ring topology or for systems in which every single transition is visible. 
    
    Furthermore, we also show that the visible entropy production rate {\color{red}can be written as the sum of} two independent contributions
    \begin{equation}
    \sigma_{\mathcal{L}} =  \sigma_\ell + \sigma_t,
    \end{equation}
    {\color{red} both of which are positive because they take the form of } Kullback-Leibler divergences of transition statistics, i.e. $\sigma_\ell\geq 0$ and $\sigma_t\geq 0$. The contribution $ \sigma_\ell $ depends solely on the mere occurrence of transitions, whereas $\sigma_t$ depends on the observed inter-transition times. Analytical expressions for $\sigma_\ell$ and $\sigma_t$ can be found in Eqs.~\eqref{gen_epr_ell} and \eqref{gen_epr_t}, copied here for convenience:
    
    \begin{eqnarray}
        \sigma_\ell  &=& \traff \sum_{\ell, \ell' \in \mathcal{L}} P(\ell \vert \ell') P(\ell') \ln \frac{P(\ell \vert \ell')} {P(\overline{\ell'}\vert \overline{\ell})}, \label{7}\\
        \sigma_t &= & \traff \sum_{\ell, \ell' \in \mathcal{L}} P(\ell \vert \ell') P(\ell') \nonumber\\
        && \phantom{\traff \sum_{\ell, \ell' \in \mathcal{L}}} \times D\left[ P(t\vert \ell', \ell) \vert \vert P(t\vert \overline{\ell}, \overline{\ell'}) \right],\label{8}
    \end{eqnarray}
    where \(\traff\) is the visible traffic rate, {\color{red}i.e. the expected number of visible transitions that occur over time~\cite{baiesi2009nonequilibrium}, sometimes also called dynamical activity~\cite{PhysRevLett.98.195702}. The sums in Eqs.~(\ref{7}-\ref{8}) run over the set of visible transitions $\mathcal{L}$ and the bar in $\overline{\ell}$ denotes the opposite direction of \(\ell\), i.e. the observed transition when the dynamics is time reversed. In Eq.~\eqref{8} and in the following, we denote by \(P(t\vert \ell', \ell)\)  the probability density for the inter-transition time between \(\ell'\) followed by \(\ell\), and we have also introduced a key quantity given by the Kullback-Leibler divergence between inter-transition time distributions
    \begin{equation}\label{eq:KLDITTgeneral}
       \hspace{0.5cm} D\left[ P(t\vert \ell', \ell) \vert \vert P(t\vert \overline{\ell}, \overline{\ell'}) \right] = \int_0^{\infty} {\rm d}t  P(t\vert \ell', \ell) \ln \frac{ P(t\vert \ell', \ell)}{P(t\vert \overline{\ell}, \overline{\ell'}) }.
    \end{equation}
    }
    
    For the relevant case of only two visible transitions in forward ("+") and backward ("-") directions between the same pair of states, i.e. $\mathcal{L}=\{ \transR,\transL\}$, Eqs.~(\ref{7}-\ref{8}) simplify to Eqs.~\eqref{epr_alpha} and \eqref{epr_t}, copied here for convenience:
    \begin{eqnarray}
        \sigma_\ell  &=& \traff [P(\transR) - P(\transL)] \ln \frac{P(\transR \vert \transR)}{P(\transL \vert \transL)},\\
        \sigma_t &= & \traff P(\transR\vert\transR) P(\transR) D [ P(t\vert \transR,\transR) \vert \vert  P(t\vert \transL,\transL)]  \nonumber\\
        &&+ \traff P(\transL\vert\transL)P(\transL) D [P(t\vert \transL,\transL) \vert \vert  P(t\vert \transR,\transR)]. \phantom{bla}
    \end{eqnarray}
    Interestingly, both depend on the statistics of {\em repeated} transitions; $\sigma_\ell$ depends on the conditional probabilities $P(\ell | \ell)$, and $\sigma_t$ depends on inter-transition time probability densities $P(t\vert \ell, \ell)$ {\color{red} through the Kullback-Leibler divergences
     \begin{eqnarray}\label{eq:KLDITT+-}
       \hspace{0.5cm} D\left[ P(t\vert +  , +) \vert \vert P(t\vert -  , -) \right] &=& \int_0^{\infty} {\rm d}t  P(t\vert +, +) \ln \frac{ P(t\vert +, +)}{P(t\vert -, -) }\nonumber\\
        \hspace{0.5cm} D\left[ P(t\vert -, -) \vert \vert P(t\vert + , +) \right] &=& \int_0^{\infty} {\rm d}t  P(t\vert -, -) \ln \frac{ P(t\vert -, -)}{P(t\vert +, +) }.\nonumber\\
    \end{eqnarray}
    }The value of \(\sigma_\mathcal{L}\) allows to improve on entropy production rate estimates previously proposed \cite{Bisker_2017, martinez2019inferring}. We also show that our approach provides a tighter bound for \(\sigma\)
    than some  of the  so-called thermodynamic uncertainty relations~\cite{barato2015thermodynamic, gingrich2016dissipation} especially in situations where the net current is small (e.g. for molecular motors close to stall force, the force at which the motor stops moving).

    \item[III)] {\it Application of the formalism to three distinct stochastic models in cell biology}: motion of {\color{red}dynein} and kinesin on linear tracks, and template-directed polymerization processes in the presence of 
    disorder. Particularly interesting from these examples is the finding that inter-transition times of  repeated {\color{red} \(P(t \vert \ell, \ell)\)} and alternate transitions {\color{red}(viz. \(P(t \vert \ell, \overline{\ell})\))} carry different information about the hidden Markov chain.  Whereas repeated transitions allow to estimate dissipation, alternated transitions provide hints about disorder.

\end{itemize}

The paper is structured as follows: in Sec.~\ref{sec:trans_stat} we develop our framework and derive Eq.~ \eqref{resultintro} for generic Markov chains; in Sec.~\ref{sec:ring} we obtain the results for a pair of transitions in opposite directions along a system with ring topology; in Sec.~\ref{sec:irr} we consider transitions over a pair of states to address the problem of estimation of  entropy production; in Sec.~\ref{sec:bio} we discuss biophysical applications for dynein, kinesin and motion in disordered tracks; finally we conclude with a discussion in Sec.~\ref{sec:discussion}. Detailed mathematical  proofs   are given in the Appendices. {\color{red}Results similar to those in the present manuscript are discussed in the coetaneous article~\cite{seifertarxiv}, see Sec.~\ref{sec:discussion} for a more detailed discussion.}

%
\section{Visible transitions' statistics}\label{sec:trans_stat}
%

\subsection{Framework}
\label{sec:framework}

We consider continuous-time Markov chains over a finite and discrete state space \(\{1,2,\ldots,N\}\). We assume that the state space structure  is such that any two states are connected by only one transition, and that the network of states is  irreducible. Thus we assume that there always exists a non-zero probability path from any to every state. The Perron-Frobenius theorem ensures the existence of a unique stationary distribution towards which the system {\color{red}relaxes and the system's ergodicity, the equivalence between time and ensemble averages}. The occupation probability at time \(t\) is expressed as a column vector $\ket{p (t)}=(p_1(t) ,p_2(t) ,\ldots ,p_N(t))^{\mathsf{T}}$ obeying the master equation
\begin{equation}\label{eq:meq}
    \dv{t} \ket{p(t)} = \mathbf{W} \ket{p(t)},
\end{equation}
where \(\mathbf{W}\) is a time-independent stochastic matrix with positive non-diagonal elements \( W_{ij}\), which are the transition rates from state \(j\) to \(i\), and negative diagonal elements \( W_{ii} = - \sum_{j\neq i} W_{ij} \) {\color{red}the escape rate from state~\(i\).}

An observer unambiguously detects transitions that belong to a subset \(\mathcal{L}\) of all possible transitions, while the remaining transitions and the occupancy of internal states go unnoticed. Visible transitions \(\ell\in\mathcal{L}\) connect state \(\lbra{\ell}\) to a different state \(\lket{\ell}\). In jump processes, transitions are instantaneous and the system spends time in states, called sojourn times. We define the inter-transition time as the sum of all sojourn times between two consecutive visible transitions.

We introduce the \textit{survival matrix} \(\mathbf{S}\), obtained by subtracting from the stochastic matrix the transition rates related to every visible transition:
\begin{equation}
    \mathbf{S} \equiv \mathbf{W} - \sum_{\ell \in \mathcal{L}} \lket{\ell} \lbra{\ell} \mathbf{W}^\mathsf{T} \lket{\ell} \lbra{\ell},
    \label{eq:Sdef}
\end{equation}
where {\color{red}the} term being summed is a matrix with all zero entries but for term \(\lbra{\ell} \mathbf{W}^\mathsf{T} \lket{\ell}\), which is the rate of transition \(\ell\).

\subsection{Main results}
\label{sec:mainres}

The survival propagator \(\exp(t\mathbf{S})\) describes {\color{red}the system's evolution given that no visible transition occurs}. It does not conserve probability because not every column {\color{red}of \(\mathbf{S}\)} adds up to zero, thus it can be interpreted as a transition matrix of a process with probability leakages whenever a transition in \(\mathcal{L}\) takes place.

Consider a succession of transitions and inter-transition times, as in Eq.\,(\ref{trajectory}), and create the histogram of {\color{red}times conditioned on the occurrence of the previous and next transitions}. This provides the empirical definition of the inter-transition times frequency:
\begin{equation}\label{timefreq}
    F_\tau(t\vert \ell_i, \ell_{i+1}) dt \sim \mathrm{histogram} (t\vert \ell_i, \ell_{i+1}).
\end{equation}
The frequency that the next observed transition is \(\ell_{i+1}\) given that the previous is \(\ell_i\) can be obtained as 
\begin{equation}\label{condfreq}
    F_\tau (\ell_{i+1}\vert \ell_i) \equiv \frac{\#(\ell_{i} \to \ell_{i+1})}{ \sum_{j=1}^{\abs{\mathcal{L}}} \#(\ell_i \to \ell_j)},
\end{equation}
where \(\#(\ell_i \to \ell_{i+1})\) is the number of transitions \(\ell_i\) followed by \(\ell_{i+1}\) and \(\abs{\mathcal{L}}\) is the number of visible transitions, the cardinality of subset \(\mathcal{L}\). Furthermore, the frequency that one observed transition is \(\ell_i\) among all transitions in a trajectory is
\begin{equation}\label{uncondfreq}
    F_\tau(\ell_i) \coloneqq \frac{\# \ell_i}{ \sum_{j=1}^{\abs{\mathcal{L}}} \# \ell_j}.
\end{equation}    
Due to the system's ergodicity, all empirical probabilities have as {\color{red}both} expected and asymptotic values the real probability, \(P(\cdot) = \langle F_\tau(\cdot)\rangle = \lim_{\tau\to\infty} F_\tau (\cdot)\).

Finding the probability that by time \(t\) the system has not performed any transitions in \(\mathcal{L}\) and then performs \(\ell\) is a {\color{red}``first-transition time''} problem whose solution leads to our main result below.

\textbf{Result:} Let \(\mathbf{W}\) be the transition matrix of a continuous-time and {\color{red}stationary} discrete-state space irreducible Markov chain, consider a subset \(\mathcal{L}\) of all possible transitions and the survival matrix as in Eq.~ \eqref{eq:Sdef}. The joint probability that the inter-transition time falls within \([t, t+dt)\) and that the next visible transition is \(\ell_{i+1}\in \mathcal{L}\), given that the last observed transition was \(\ell_{i} \in \mathcal{L}\), is
\begin{equation}\label{theorem}
    P(t, \ell_{i+1} \vert \ell_i) dt = \lbra{\ell_{i+1}} \mathbf{W}^\mathsf{T} \lket{\ell_{i+1}} \lbra{\ell_{i+1}}\exp(t\mathbf{S})\lket{\ell_i} dt,
\end{equation}
in agreement with \cite{seifertarxiv}. See  Appendix \ref{proof} for a proof. 

All other probabilities we are interested in can be  obtained from Eq.~\eqref{theorem}, whose joint probability can be split into \(P(t,\ell_{i+1} \vert \ell_i ) = P(\ell_{i+1} \vert \ell_i ) P(t\vert \ell_i, \ell_{i+1})\). The conditional probability of the next observed transition can be obtained by integrating over time, resulting in
\begin{align}\label{transprob}
    P(\ell_{i+1}\vert \ell_i) &= \int\limits_0^\infty \mathrm{d}t P(\ell_{i+1}, t\vert \ell_{i}) \nonumber\\
    &=- \lbra{\ell_{i+1}} \mathbf{W}^\mathsf{T} \lket{\ell_{i+1}} \lbra{\ell_{i+1}} \mathbf{S}^{-1} \lket{\ell_{i}}.
\end{align}
Without {\color{red}the} need for additional assumptions, the probability density of inter-transition time \(t\) between such transition and its preceding one can be obtained by dividing the joint probability by the transition probability above,
\begin{equation}\label{eq:dwell}
    P (t\vert \ell_{i}, \ell_{i+1}) = \frac{P(t,\ell_{i+1} \vert \ell_{i})}{P(\ell_{i+1} \vert \ell_{i})} = - \frac{ \lbra{ \ell_{i+1}} \exp(t \mathbf{S}) \lket{ \ell_{i} }}{ \lbra{ \ell_{i+1}} \mathbf{S}^{-1} \lket{ \ell_{i}} }.
\end{equation}
{\color{red}It} satisfies \(\int_0^\infty dt P (t\vert \ell_{i}, \ell_{i+1}) = 1\) for all \(\ell_i\) and \(\ell_{i+1}\).

\begin{figure*}[ht!]
    \centering
    \includegraphics[width=0.95\textwidth]{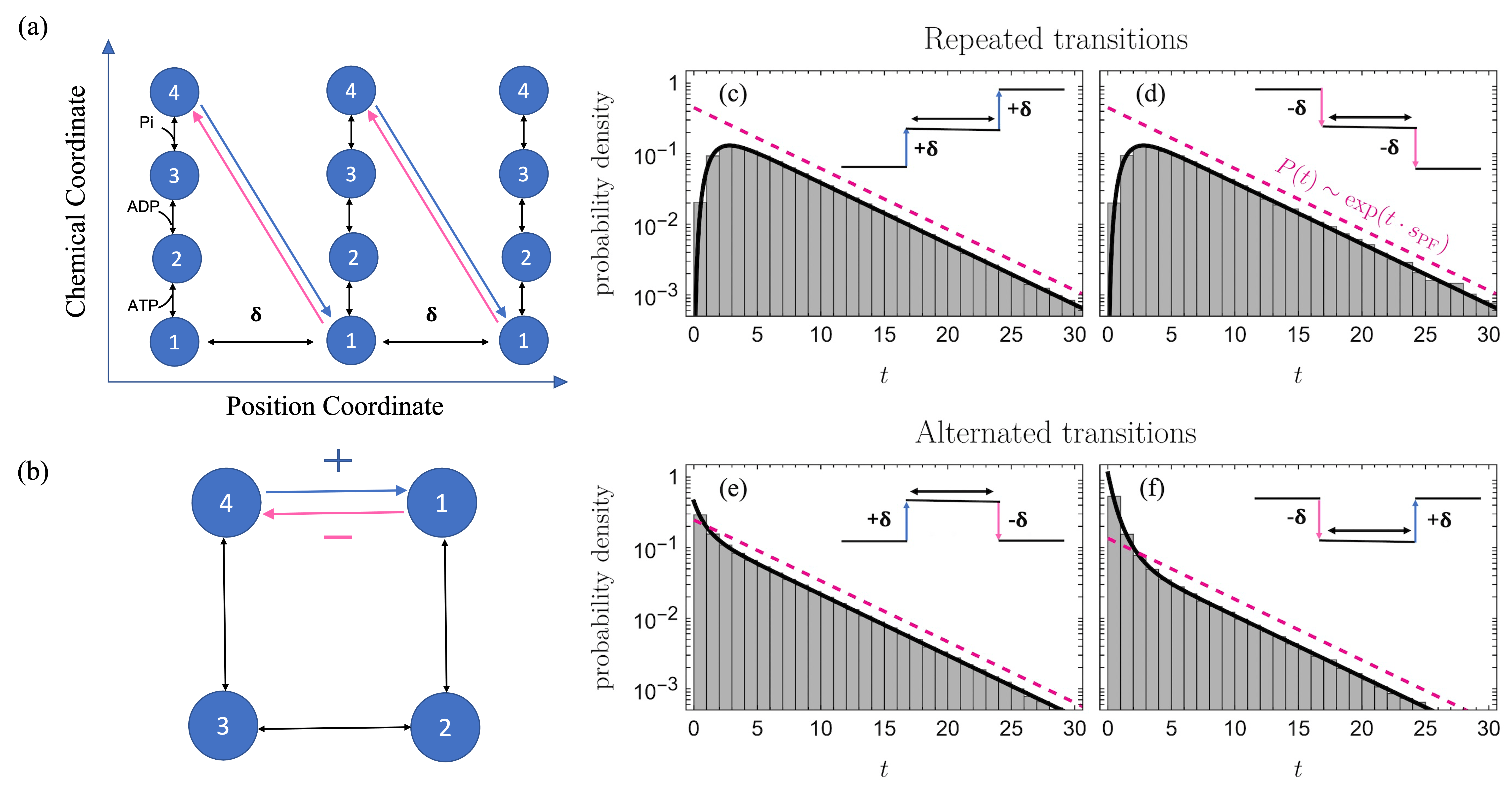}
    \caption{Illustration of transition statistics in an ATP-driven motion of a molecular machine in which only spatial displacements are visible to an experimental apparatus. Chemical transitions (ATP binding $1\to 2$, ADP release $2\to3$ and Pi release $3\to4$) are hidden to the apparatus.  (a,b) Illustration of the model state network, given by a four-state model (b) arranged along a periodic spatial lattice (a) along which the motor moves only when executing the $4\to 1$ transition (motion forward \(+\) with step size  $\delta$) or the $1\to 4$ transition  (motion backward \(-\)  with step size  $\delta$). (c)-(f) Histograms for the inter-transition time probability densities obtained from Gillespie simulations (bars),  and analytical predictions given by Eq.~\eqref{eq:dwell}. The dashed magenta lines have slope given by the largest eigenvalue of the survival operator \(\mathbf{S}\), {\color{red} see Eq.~\eqref{tail}}, and the inserted diagrams are illustration of the displacement of the motor during the respective pair of transitions (see insets).}
    \label{fig:summary_pic}
\end{figure*}

We name the expected number of visible transitions over time \(\traff\) the \textit{visible traffic rate}, inspired by  time-symmetric quantities relevant in the analysis of stochastic systems far from equlibrium~\cite{MAES20201}, sometimes referred to as dynamical activity \cite{PhysRevLett.98.195702} and frenesy \cite{baiesi2009nonequilibrium,PhysRevE.100.042108,maes2019nonequilibrium}. Analytically its value can be obtained in the limit $t \to \infty$ as 
\begin{equation}
    \traff = \sum_{\ell \in \mathcal{L}} \lbra{\ell} \mathbf{W}^\mathsf{T} \lket{\ell} \lbra{\ell}p_\infty\rangle ,
\end{equation}
where $\ket{p_\infty} $ is the stationary distribution given by the solution of  $ \mathbf{W} \ket{p_\infty} = 0$.
Furthermore, the stationary probability that a visible transition is \(\ell \in \mathcal{L}\) is given by
\begin{equation}\label{uniquetrans}
    P(\ell) = \frac{1}{\traff} \lbra{\ell} \mathbf{W}^\mathsf{T} \lket{\ell} \lbra{\ell} p_\infty \rangle.
\end{equation}

{\color{red}
\subsection{Numerical illustration of the framework}
}

Figure~\ref{fig:summary_pic} presents an example of the application of our approach to a model of a  molecular motor with four internal states that {\color{red}are} driven by the consumption of adenosine triphosphate (ATP)~\cite{chemla2008exact}. The motor performs  spatial displacements along a filament along the only visible transition in a single-molecule experiment. Fig. \ref{fig:summary_pic}(a) is a scheme of the motor's motion, transitions in the chemical coordinate involve consumption and production of chemical species and are considered as invisible for the experimenter, conversely transitions in the position coordinate \(1\leftrightarrow 4\) are considered as visible since they result in spatial displacement of size \(\delta\) (mechanical movement), in this case they compose the subset \(\mathcal{L}\); (b) shows the irreducible network in which a Markov chain describes the evolution and visible transitions \(4\to 1 \equiv \transR\) and \(1\to 4 \equiv \transL\) are respectively related to forward and backwards displacement; (c)-(f) shows an excellent agreement between numerical simulations and  Eq.~\eqref{eq:dwell} for all the  distributions of inter-transition times between repeated (\(\transR \transR\),\(\transL\transL\)) and alternated (\(\transR\transL\), \(\transL\transR\)) transitions.  While alternated transitions yield an inter-transition time distribution that is monotonously decreasing, the distribution of inter-transitions times between repeated transitions  is non-monotonous. This is because  of network topology constraints: whereas for alternated transitions $t=0$ is the most likely event, repeated transitions require motion over the entire hidden network, which renders the probability of $t=0$ almost impossible for large hidden networks. Furthermore, we observe that the distributions of all inter-transition times have the same exponential tail (magenta dashed line in Fig.~\ref{fig:summary_pic} c-f). This is consistent with theory, as discussed in the next Sec.~\ref{sec:remarks}.
 
\subsection{Additional remarks}
\label{sec:remarks}

(i) {\em Moments of the inter-transition times.} 
Equation~ \eqref{eq:dwell} is key for further results of this work. An immediate outcome is that, since Eq.~ \eqref{eq:dwell} is the {\color{red} probability density of the inter-transition time between} of \(\ell_{i+1}\) after \(\ell_{i}\), the mean inter-transition time can also be obtained from the survival matrix:
\begin{equation}
    \int_0^\infty \dd{t}  t P (t\vert \ell_{i}, \ell_{i+1}) =  - \frac{ \lbra{ \ell_{i+1}} (\mathbf{S}^{-1})^2 \lket{ \ell_{i} }}{ \lbra{ \ell_{i+1}} \mathbf{S}^{-1} \lket{ \ell_{i}} }.
\end{equation}
and higher-order moments can be obtained analogously. 

(ii) {\em Generalization of first-passage times.} First-passage times between states  can be obtained as a particular case of {\color{red}first-transition} times. Let \(\mathcal{L}_i = \{i'\to i:\forall i'\neq i\}\) be the set of all transitions leading to an absorbing state \(i\). The first time that state \(i\) is reached coincides with the first time that one of these transitions is observed. From Eq.~\eqref{theorem} we then obtain the probability density for the first-passage time of reaching state \(i\) starting from a state \(j\neq i\)
\begin{align}
    P(t,i\vert j)dt &= \sum_{\ell\in \mathcal{L}_i} P(t,\ell \vert j)dt \nonumber\\
    &= \sum_{\ell\in \mathcal{L}_i} \lbra{\ell} \mathbf{W}^\mathsf{T} \lket{\ell} \lbra{\ell} \exp(t\mathbf{S}) \ket{j}dt,
\end{align}
where \(P(t,\ell \vert j)dt\) is the probability that, starting from $j$, the first visible transition observed in the time interval $[t,t+dt)$ is $\ell$. This latter result is well-known, see e.g.~\cite{vankampen1992}.

(iii) {\em Connection to large deviation theory.}
Consider the number of times  \(\#_t(\ell)\) a transition is performed up to time \(t\) (sometimes called flux, or counting field). Notice that $\#_t(\ell)$ only vanishes for all $\ell \in \mathcal{L}$ if no visible transition has been performed. {\color{red}Therefore} the generating function of its moments \(\langle e^{\lambda {\color{red}\sum_{\ell\in\mathcal{L}} \#_t(\ell) }} \rangle\) in the limit \(\lambda \to - \infty\) is precisely the survival probability density. The moment generating function can be calculated as \( \sum_y \bra{y} \exp( t\mathbf{W}_\lambda) \ket{x}\) \cite{garrahan}, where \(\mathbf{W}_\lambda\) is the so-called tilted matrix, which in the limit \(\lambda \to - \infty\) reduces to $\mathbf{S}$, consistently with Eq.~\eqref{theorem}.

(iv) {\em Existence of \(\mathbf{S}^{-1}\).}
Since the process defined by \(\mathbf{W}\) is ergodic, for a large enough time the system will perform at least one of the observed transitions with probability one: \(\lim_{t\to\infty} \ket{p(t)} = \lim_{t\to\infty} \exp(t\mathbf{S}) \ket{p(0)} = \vec{0}\), where \(\vec{0}\) is a vector of zeroes. This is ensured by the fact that every eigenvalue of \(\mathbf{S}\) has a negative real part, \(\mathrm{Re}(\lambda_i) < 0\ \forall i\){\color{red}, as proved in the Supplementary Material of Ref.~\cite{harunari2022beat}}. Such property also guarantees the convergence of the integral \(\int_0^\infty \dd{t} \exp(t\mathbf{S})\), which is required to normalize the probability in Eq.~ \eqref{eq:dwell}, and \(\mathrm{det}(\mathbf{S}) = \prod_i \lambda_i \neq 0\), which grants the existence of \(\mathbf{S}^{-1}\).

(v) {\em Probability of instantaneous pairs.}
The propagator acting over a state results in a probability vector with non-negative entries, \(\exp(t\mathbf{S})\lket{\ell_i}\geq 0\), therefore \(\partial_t \lbra{\ell_{i+1}}\exp(t\mathbf{S}) \lket{\ell_{i}} = \lbra{\ell_{i+1}} \mathbf{S} \exp(t\mathbf{S}) \lket{\ell_{i}}\) has the same sign as \(\lbra{\ell_{i+1}} \mathbf{S} \lket{\ell_{i}}\) at \(t=0\). If the transition \(\ell_{i+1}\) starts in the same state where \(\ell_i\) ended, \(\lbra{\ell_{i+1}} = \lket{\ell_{i}}\), the inter-transition time has non-vanishing probability of being zero since the diagonal entries of \(\partial_t \lbra{\ell_{i+1}}\exp(t\mathbf{S}) \lket{\ell_{i}}\) are always negative. Conversely, for sequences of transitions with \(\lbra{\ell_{i+1}} \neq \lket{\ell_{i}}\) the null inter-transition time has zero probability: the observer has to wait for internal jumps to occur before \(\ell_{i+1}\) takes place. This property explains the shape of inter-transition time probability densities in Fig. \ref{fig:summary_pic}: for alternated transitions \(+-\) and \(-+\) the source state of the second transition is the target of the first transition, {\color{red}therefore the probability of instantaneous inter-transition time is non-zero}  [cf. panels (d) and (e)]. On the other hand, for repeated transitions \(++\) and \(--\),  instantaneous  inter-transition times cannot be realized because one needs to perform additional transitions   [cf. panels (c) and (f)].

(vi) {\em Universality of the tails.}
{\color{red}Notice} that it is always possible to decompose the numerator in Eq.~\eqref{eq:dwell} as \(\lbra{\ell_{i+1}} \exp(t\mathbf{S}) \lket{\ell_{i}} = \sum_{k=1}^N c_k e^{t s_k}\), where \(s_k\) are the eigenvalues of \(\mathbf{S}\) and \(c_k\) are real coefficients obtained by projecting onto its eigenvectors, under the assumption that $\mathbf{S}$ has a non-degenerate spectrum, and with minor modifications of the argument otherwise \cite{polettini2014fisher}. {\color{red}Assuming \textit{hidden irreducibility}, i.e. the irreducibility of the state space after the removal of all visible transitions,} this property implies that the long-time behavior of the inter-transition time distribution is independent of the visible transitions \(\ell_{i+1}\) and \(\ell_{i}\):
\begin{equation}\label{tail}
    \lim_{t\to\infty} \frac{1}{t} \ln \left[ \lbra{\ell_{i+1}} \exp(t\mathbf{S}) \lket{\ell_{i}} \right] \asymp s_{\mathrm{PF}}.
\end{equation}
where $s_{\mathrm{PF}}$ is the dominant Perron-Froebenius root{\color{red}, a negative real value.} Therefore, all inter-transition time distributions have the same exponential tail given by the largest eigenvalue of~\(\mathbf{S}\).
{\color{red}This can be observed in Figs.~\ref{fig:summary_pic}(c)-(f), where tails of the histograms obtained for the four types inter-transition times match the value given by  \(s_\mathrm{PF}\).}

%
\section{Explicit results for unicyclic networks}\label{sec:ring}
%

{\color{red}In addition to the developed generic framework, analytical expressions for inter-transition time distributions in ring (unicyclic) networks can be obtained  using Laplace transforms. To this aim, we now use a combinatoric graph-theoretic approach based on sums over all possible hidden paths in Laplace space. As shown below, 
these explicit calculations showcase that computing inter-transition statistics in generic Markov chains is often a Herculean task, which is greatly simplified by the exact analytical framework developed  in Sec.~\ref{sec:trans_stat}. }

In a variety of models of e.g. enzymatic reactions~\cite{qian2006generalized} the state space can be depicted as a ring network, where every state \(i\) is connected to only its nearest neighbours \(i\pm 1\) and nothing else (\(1 \leftrightarrow 2 \leftrightarrow \ldots \leftrightarrow N \leftrightarrow 1\)). In particular we consider as visible the pair of transitions between states \(1\leftrightarrow N\), without loss of generality. For this section we also assume that every neighboring states have transitions in both directions, \(W_{i,i+1}>0\) and \(W_{i+1,i}>0\), allowing for the cycle performance in both orientations.

We denote the two visible transitions as follows, the clockwise transition from state \(N\) to \(1\) is \(\transR \equiv N\to 1\), and the counterclockwise transition is \(\transL \equiv 1\to N\). There are four possible inter-transition times to be considered, {\color{red}between pairs of successive transitions} \(\transR\transR\), \(\transR\transL\), \(\transL\transR\), and \(\transL\transL\).

The probability density of spending time \(\tau_j\) in a given state \(j\) before the next transition to \(i\) (often called sojourn time) is given by
\begin{equation}
    \pi_{ij}(\tau_j) \coloneqq P(\tau_j, j\to i \vert j) = W_{ij} \exp(-W_{jj} \tau_j).
\end{equation}
To characterize the different paths that intertwine the desired transitions, we introduce the number of times the pair of opposite transitions between \(i\leftrightarrow i+1\) are performed as \(k_i\) and the number of possible paths satisfying \(\vec{k} = (k_1,\ldots , k_{N-1}, k_N=0)\) is \(C^{\vec{k}}_{\transR,\transR}\). The simplest ``bare'' path leading to \(\transR\transR\) is a sequence of clockwise transitions starting and ending in 1, and the probability density of performing it in an interval \(t\), up to a normalization constant, is given by the convolution of sojourn times
\begin{equation}
    P_\text{bare} (t\vert +,+) \propto \int \dd{\vec{\tau}} \pi_{1,N}(\tau_N) \cdots \pi_{2,1}(\tau_1) \delta\left( t -\sum_i \tau_i \right),
\end{equation}
where \(\delta\) is the Dirac delta distribution. As the Laplace transform of convolutions is the product of Laplace transforms we further deal with products of terms in the form of
\begin{equation}
    \widehat{\pi}_{ij}(s) \equiv \int_0^\infty \dd{\tau_j} \pi_{ij}(\tau_j) e^{-s\tau_j} = \frac{W_{ij}}{W_{jj} + s},
\end{equation}
where the hat \(\widehat{\cdot}\) denotes the Laplace transform and for simplicity we often suppress the dependency on \(s\), the complex frequency corresponding to time in the Laplace space.

{\color{red}To count all trajectories we solve a non-trivial combinatoric problem introducing the concept of {\em backbone}: for a given path, its associated backbone is composed by  the set of every last transition performed between each pair of visited states. Once the backbone associated with a path is identified, all other variables in the trajectory can freely change without changing the fact that the trajectory starts and ends at  two prescribed visible transitions.}
For example, in the case of repeated transitions \(++\) a path characterized by \(\vec{k}\) will contain \(k_1\) transitions \(2\to 1\) and \(k_1+1\) transitions \(1\to 2\). This last performed transition \(1\to 2\) ensures that the path is moving in the direction of eventually performing \(\transR\) again and is part of the backbone; the rest  of the backbone will come from transitions \(2\to 3\), \(3\to 4\) and so on.

\begin{figure}[t!]
    \centering
    \includegraphics[width=.5\columnwidth]{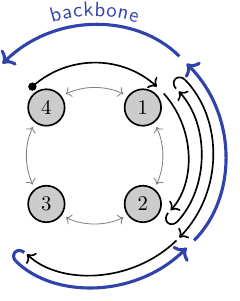}
    \caption{{\color{red} Illustration of a single trajectory in a four-state Markov process where only transitions \(4\to 1\) (+) and \(1\to 4\) (-) are visible. Thin gray arrows represent the possible transitions in the state space, while thick arrows represent the trajectory. For this example, we define its associated "backbone" (blue thick arrows) by the sequence of the transitions last travelled between each pair of states that were visited in the trajectory:  \(3\to 2\),  \(2\to 1\) and  \(1\to 4\), see text for further details.}}
    \label{fig:backbones++}
\end{figure}

\subsection{Repeated transitions}\label{repeated_ring}

For the case \(\transR\transR\), notice that each pair of states \(i\leftrightarrow i+1\) accommodates \(2k_i\) transitions in a path, half clockwise and half counterclockwise, and then one extra transition that belongs to the backbone, ensuring that the path is not stuck between these two states. The inter-transition time probability density can be obtained from the convolution of every sojourn time in a path, and by summing over all possible paths. Its Laplace transform is given by
\begin{align}\label{eq:PRRC}
    \widehat{P}(s\vert \transR,\transR) =& \frac{1}{\mathcal{N}_{\transR, \transR}} \overbrace{\left( \prod_{i=1}^{N} \widehat{\pi}_{i+1,i} \right)}^{\text{backbone}} \times \nonumber\\ &\sum_{k_1,\ldots,k_{N-1}=0}^\infty C_{\transR,\transR}^{\vec{k}} \prod_{i=1}^{N-1} [ \underbrace{ \widehat{\pi}_{i,i+1} \widehat{\pi}_{i+1,i}}_{\equiv x_i} ]^{k_i},
\end{align}
where we define \(x_i = \widehat{\pi}_{i,i+1} \widehat{\pi}_{i+1,i}\) as the product of Laplace transformed sojourn times in two opposite directions {\color{red}and we impose the condition} \(N+1 \equiv 1\).

{\color{red}The initial value theorem for Laplace transforms states that \(P(0) = \lim_{s\to\infty} s \widehat{P}(s)\), theferore} the constant \(\mathcal{N}_{\transR, \transR}\) can be obtained by \(\widehat{P}(0\vert \transR,\transR) = 1\), which ensures that the inverse Laplace transform \(P(t\vert\transR,\transR)\) is normalized. The combinatorial coefficient \(C_{\transR,\transR}^{\vec{k}}\) is the number of all possible paths between \(\transR\transR\) with transitions satisfying \(\vec{k}\). It is obtained in Appendix \ref{sec:appcombinatorics} and reads
\begin{equation}\label{eq:combinatoric}
    C_{\transR,\transR}^{\vec{k}} = \prod_{i=2}^{N-1} \binom{k_i+k_{i-1}}{k_{i}}.
\end{equation}

Eq.~ \eqref{eq:PRRC} simplifies by plugging {\color{red}in} Eq.~ \eqref{eq:combinatoric} and introducing a {\color{red}continued} fractions generator $\Theta[x_i] \coloneqq x_i/(1-\Theta[x_{i-1}])$ that truncates at $\Theta[x_1] = x_1$. From the property $\sum_{k=0}^\infty \binom{n+k}{k} x^k = (1-x)^{-(n+1)}$, valid for \(\abs{x}<1\), we obtain a simplified expression
\begin{equation}\label{eq:PRRlaplace}
    \widehat{P}(s\vert \transR,\transR) = \frac{1}{\mathcal{N}_{\transR, \transR}} \widehat{\pi}_{1,N} \prod_{i=1}^{N-1} \frac{ \widehat{\pi}_{i+1,i}}{1-\Theta[x_i]}.
\end{equation}

The case \(\transL\transL\) can be obtained analogously upon the substitutions $i\mapsto N-i+1$, $\forall i\in [1,N]$ and \(\Xi[x_i] \coloneqq x_i/(1-\Xi[x_{i+1}])\) with \(\Xi[x_{N-1}] = x_{N-1}\):
\begin{equation}\label{eq:PLLlaplace}
    \widehat{P}(s\vert \transL,\transL) = \frac{1}{\mathcal{N}_{\transL, \transL}}  \widehat{\pi}_{N,1} \prod_{i=1}^{N-1} \frac{ \widehat{\pi}_{i,i+1}}{1-\Xi[x_{i}]}.
\end{equation}

By a diagrammatic approach to obtain explicitly the {\color{red}continued} fraction generators \(\Theta\) and \(\Xi\) (Appendix~\ref{diagram}) we find that inter-transition times densities are the same for repeated transitions 
\begin{equation}\label{eq:haldane}
    P(t\vert \transR,\transR) = P(t\vert \transL,\transL).
\end{equation}
Such property is reminiscent of the so-called generalized Haldane equality \cite{qian2006generalized, ge2008waiting, PhysRevX.7.011019}, which states that the probability density of waiting time \(t\) until a system performs a clockwise cycle, given that a counterclockwise cycle was not performed, is the same of waiting \(t\) for the opposite phenomenon. This property can be observed in Figs.~ \ref{fig:summary_pic}(d) and (e), and in general is not satisfied for alternated transitions.

 Since \(\lim_{s\to\infty} \widehat{\pi}_{ij} = 0\) for every pair \(i\), \(j\), applying the initial value theorem to Eqs.~\eqref{eq:PRRlaplace} and \eqref{eq:PLLlaplace} results in
\begin{equation}\label{initialvalue}
    P(0\vert \transR, \transR) = P(0\vert \transL, \transL) = 0,
\end{equation}
apart from very specific choices of transition rates that might forbid the existence of the limit. This result confirms that instantaneously performing a full cycle has zero probability and {\color{red}gives a characteristic shape to} the histograms in Fig.~\ref{fig:summary_pic}(c) and (f).

\subsection{Alternated transitions}

\begin{figure}[t!]
    \centering
    \includegraphics[width=.8\columnwidth]{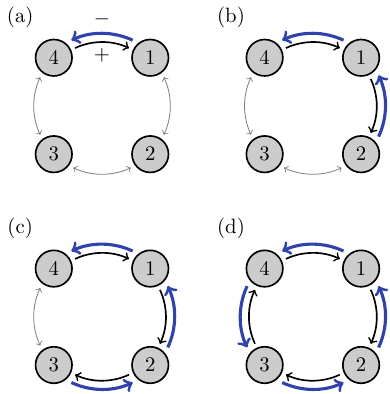}
    \caption{{\color{red}Illustration of the four possible backbones (thick blue arrows) that ensure the completion of a (-) \(1\to 4\) transition after a (+) \(4\to 1\) transition, for the example of a 4-state Markov model shown in Fig.~\ref{fig:backbones++}. Thick gray arrows represent the possible transitions, while thick arrows represent the transitions used by trajectories of different lengths. From panels (a) to (d), the trajectories have \(M=0\) to \(M=3\). More detail in the main text.}}
    \label{fig:backbones}
\end{figure}

{\color{red}Between} alternated transitions it is not necessary to cover the whole state space. In fact, it is possible to not have any transitions in between the visible ones and this is how a zero inter-transition time might occur, thus there is no analogue of Eq.~\eqref{initialvalue} for alternated transitions.

In this case there are \(N\) possible backbones, they are composed of \(M+1\in [1,N]\) transitions with the same orientation starting from the farthest visited state to the target of the last visible transition, see Fig.~\ref{fig:backbones}. Also, an even number \(2k_i\) of transitions take place between pairs \(i\leftrightarrow i+1\).

The Laplace transform of the inter-transition time probability density for the pair \(\transR\transL\) is
\begin{align}
    \widehat{P}(s\vert \transR,\transL) =& \frac{1}{\mathcal{N}_{\transR, \transL}} \widehat{\pi}_{N,1}\times \nonumber\\ 
    &\sum_{M=0}^{N-1} \sum_{k_1,\ldots ,k_M = 1}^\infty C_{\transR,\transL}^{\vec{k},M} \prod_{i=1}^{M}  [ \underbrace{ \widehat{\pi}_{i,i+1} \widehat{\pi}_{i+1,i}}_{\equiv x_i} ] ^{k_n},
\end{align}
where the backbone contributions come from the sum over \(M\) and \( \widehat{\pi}_{N,1} \prod_{i=1}^M \widehat{\pi}_{i,i+1}\).

The coefficient
\begin{equation}
    C_{\transR,\transL}^{\vec{k},M} = \prod_{i=2}^{M} \binom{k_i+k_{i-1}-1}{k_i } 
\end{equation}
counts the number of possible paths leading to \(\transR\transL\) with a given \(\vec{k}\) and backbone length of \(M\) (more details in Appendix \ref{sec:appcombinatorics}). Once again from the property $\sum_{k=0}^\infty \binom{n+k}{k} x^k = (1-x)^{-n-1}$ we obtain a simplified expression
\begin{equation}\label{eq:PRLlaplace}
    \widehat{P}(s\vert \transR,\transL) \propto \widehat{\pi}_{N,1} \sum_{M=0}^{N-1} \frac{\prod_{i=1}^M x_i}{\prod_{j=1}^{M-1} (1-\Xi[x_j])^2} \frac{1}{1-\Theta[x_M]}
\end{equation}
and analogously we find
 \begin{align}\label{eq:PLRlaplace}
    \widehat{P}(s\vert \transL\transR) \propto \widehat{\pi}_{1,N} \sum_{M=0}^{N-1}& \frac{\prod_{i=N-M}^{N-1} x_i}{\prod_{j=N-M+1}^{N-1} (1-\Xi(x_j))^2} \times \nonumber\\ &\frac{1}{1-\Xi(x_{N-M})}.
\end{align}

Applying the {\color{red}initial value} theorem to Eqs.~\eqref{eq:PRLlaplace} and \eqref{eq:PLRlaplace} results in
\begin{equation}\label{initialvalue2}
    P(0\vert \transR, \transL) = \frac{W_{N,1}}{\mathcal{N}_{\transR, \transL}}
\end{equation}
and
\begin{equation}\label{initialvalue3}
    P(0\vert \transL, \transR) = \frac{W_{1,N}}{\mathcal{N}_{\transL, \transR}},
\end{equation}
{\color{red}we recall that \(\mathcal{N}_{\transR, \transL}\) and \(\mathcal{N}_{\transR, \transL}\) can be obtained by \(\widehat{P}(0\vert \bullet) = 1\) as a property of Laplace transforms, since \(P(t\vert \bullet)\) is normalized.} The non-vanishing contribution comes from the terms with \(M=0\), which means that it is possible to instantaneously observe a pair of alternated transitions and it is due to the shortest backbone of all: a single transition. This can be observed in the shape of histograms in Fig.~\ref{fig:summary_pic}(d) and (e).

To obtain the inter-transition time densities one needs to perform an inverse Laplace transform on Eqs.~\eqref{eq:PRRlaplace}, \eqref{eq:PLLlaplace}, \eqref{eq:PRLlaplace} and \eqref{eq:PLRlaplace}. We remark that{\color{red}, while possible,} in general it is not straightforward to find closed analytical expressions to such inverse Laplace transforms (cf. Appendix B of \cite{Polettini_2015}). {\color{red}Notice} that it is possible to obtain all moments of inter-transition times without resorting to the inverse Laplace transforms by using the relation
\begin{equation}
    \expval{t^n\vert \ell_i, \ell_{i+1}} = (-1)^n \left[ \pdv[n]{s} \widehat{P} (s\vert \ell_i, \ell_{i+1})\right]_{s=0}.
\end{equation}

%
\section{Irreversibility and entropy production}\label{sec:irr}
%

Entropy production and time irreversibility are the thermodynamic footprints of nonequilibrium dynamics. In stochastic thermodynamics irreversibility of nonequilibrium stationary processes can be  quantified by the asymmetry between a process and its time reversed in {\color{red}terms} of the Kullback-Leibler divergence of forward to backward {\color{red}probabilities}~\cite{PhysRevE.85.031129} that provide bounds for the rate of entropy production. {\color{red}As we show now,} in a jump process this asymmetry is present in the sequence of visited states and, also, in the inter-transition times. In this section we introduce an inference scheme for the entropy production rate of a system for which only {\color{red}a} few transitions are visible. We also assume {\color{red}visible reversibility, i.e. that every visible transition can be performed in its opposite direction, and the opposite of a visible transition is also visible}. This scenario is typical in physical settings such as electron hopping between leads or a molecular motor walking along a microtubule.

The stationary rate of entropy production in the system plus environment is a measure of time-reversal asymmetry in the dynamics of the system averaged over all microscopic trajectories $\gamma_\tau$ over state space:
\begin{equation}\label{epr_kdl}
    \sigma   =  \lim_{\tau \to \infty} \frac{1}{\tau} \underbrace{\sum_{\gamma_\tau} P[\gamma_\tau] \ln \frac{P[\gamma_\tau]}{P[\overline{\gamma}_\tau]}}_{\displaystyle D\left( P[ \gamma_\tau ] ||P[ \overline{\gamma}_\tau ] \right)},
\end{equation}
where \(\overline{\gamma}_\tau\) is the time-reversed trajectory obtained by reverting in time the states visited along the trajectory~\({\gamma}_\tau\). For Markovian nonequilibrium time-independent processes, it has been shown~\cite{roldan2010estimating} that  the entropy production~\eqref{epr_kdl} depends only on the statistics of jumps between different states as follows
\begin{equation}
    \sigma = \sum_{i<j} J_{ij} \ln \frac{ W_{ij}}{W_{ji}}
\end{equation} 
where
\begin{equation}
    J_{ij} = W_{ij}p_j(\infty) - W_{ji} p_i(\infty)
\end{equation}
denotes the stationary probability current from state $j$ to state $i$~\cite{RevModPhys.48.571}, and for convenience we have set the Boltzmann constant \(k_\text{B}\) to unity. {\color{red}Currents can be empirically observed when the involved transitions are visible. In other words, if \(\ell = j \to i\), the current can be empirically obtained by
\begin{equation}
    J_{ij} = \lim_{\tau\to\infty} \frac{\#\ell - \#\overline{\ell}}{\tau},
\end{equation}
where we recall that \(\tau\) is the trajectory duration and \(\#\) represents the number of occurrences.
}

The entropy production rate from the available data in the present framework is obtained by comparing visible trajectories \(\Gamma_\tau^\mathcal{L}\) that can be seen as a transition based coarse-graining of the full trajectory \(\gamma_\tau\) over state space.  A key result for our estimates is the chain rule for the Kullback-Leibler divergence between two random variables~\cite{Cover2006} which has been applied also to stochastic processes~\cite{PhysRevE.78.011107, PhysRevE.85.031129}: $D[\rho_1(x,y)||\rho_2(x,y)]\geq D[\rho_1(x)||\rho_2(x)]$ for any two distributions $\rho_1$ and $\rho_2$ of two random variables $x$ and $y$. Because the trajectories $\Gamma^{\mathcal{L}}_\tau$ contain less random variables than the microscopic trajectories $\gamma_\tau$, e.g. most of the transitions and their associated inter-transition times are not included in $\Gamma^{\mathcal{L}}_\tau$, one gets  $D\left( P[ \gamma_\tau ] ||P[ \overline{\gamma}_\tau ]\right) \geq D\left( P[ \Gamma_\tau^{\mathcal{L}} ] ||P[ \overline{\Gamma}_\tau^{\mathcal{L}} ]\right)$,  which implies the inequality
\begin{align}\label{inf_definition}
    \sigma_{\mathcal{L}} = \lim_{\tau\to\infty} \frac{1}{\tau}D\left( P[ \Gamma_\tau^{\mathcal{L}} ] ||P[ \overline{\Gamma}_\tau^{\mathcal{L}} ]\right) \leq \sigma.
\end{align}
In the following, we will employ the transitions information using  $\sigma_{\mathcal{L}} $ to obtain lower bounds for the entropy production, and analyze how $\sigma_{\mathcal{L}} $ can be computed in practice from simulations or experimental data. 

\subsection{Inference of entropy production}

\begin{figure}[t!]
    \centering
    \includegraphics[width=0.8\columnwidth]{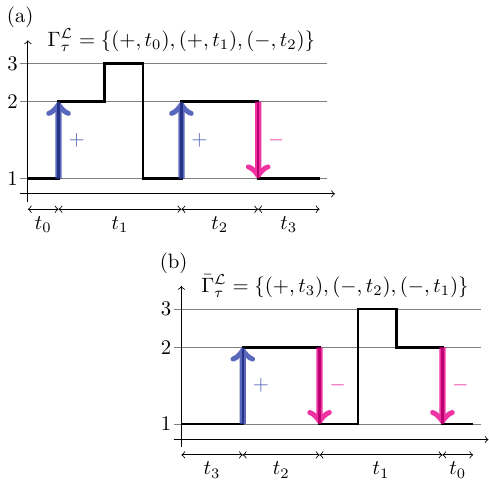}
    \caption{(a) Hidden trajectory over states \(\{1,2,3\}\) and the observation \(\Gamma_\tau^\mathcal{L}\) for \(\tau = t_0 + t_1 + t_2 + t_3\) and \(\mathcal{L} = \{ \transR : 1\to 2, \transL : 2\to 1\}\). (b) Time-reversed hidden trajectory and resulting \(\overline{\Gamma}_\tau^\mathcal{L}\).}
    \label{fig:traj_timereversal}
\end{figure}

We ask the question of how the inferred entropy production rate \(\sigma_\mathcal{L}\) can be computed in practice and study how tight the lower bound is. The observer collects a coarse-grained trajectory during an interval \([0,\tau]\) comprising visible transitions in both directions \(\ell \in \mathcal{L}\) and the inter-transition times between them:
\begin{equation}
    \Gamma_\tau^\mathcal{L} = \left\lbrace (\ell_0,t_0), (\ell_1, t_1), \ldots , (\ell_n, t_n) \right\rbrace,
\end{equation}
with \(\sum_{i=0}^{n+1} t_i = \tau\).
The construction of the time-reversed trajectory in the transition space requires special care. The time-reversed trajectory is given by the sequence of reversed transitions \(\overline{\ell}\) in the opposite order and the inter-transition times are shifted: if the time before a transition \(\ell_i\) is \(t_i\), in the time-reversed dynamics the time before transition \(\overline{\ell}_i\) is \(t_{i+1}\), see Fig.~\ref{fig:traj_timereversal} for an illustrative example. Thus
\begin{equation}
    \overline{\Gamma}_\tau = \left( (\overline{\ell}_n,t_{n+1}), (\overline{\ell}_{n-1},t_n), \ldots , (\overline{\ell}_0 , t_1) \right).
\end{equation}

\begin{figure*}[ht]
    \centering
    \includegraphics[width=.95\textwidth]{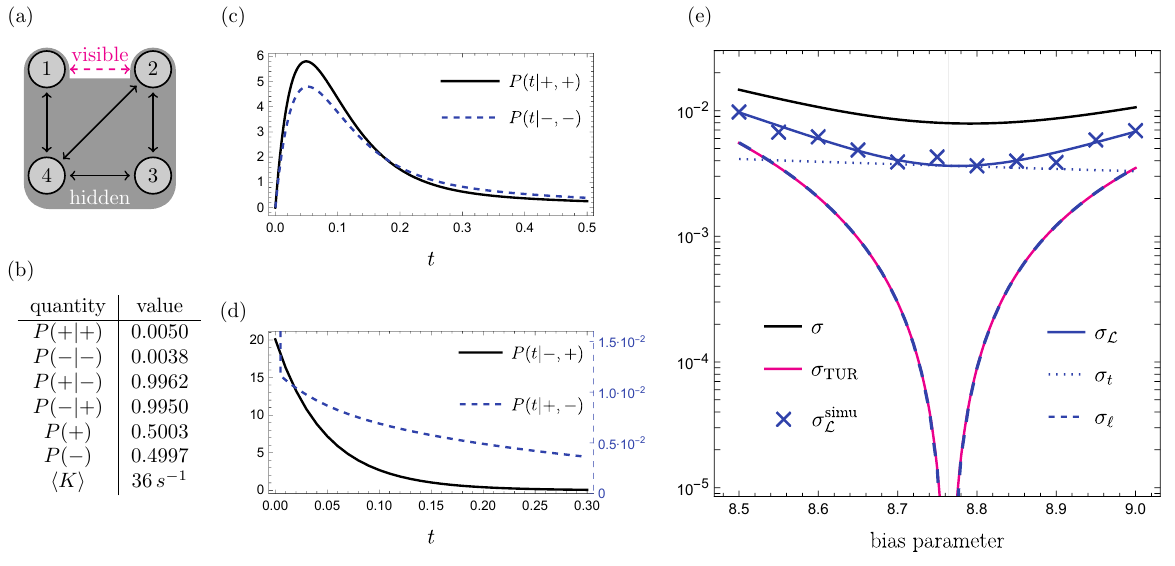}
    \vspace{-10pt}
    \caption{\textit{Estimating entropy production rate (\(k_\text{B}/s\)) from repeated transition statistics}: Illustration of the inference scheme for a network depicted in (a) with the observation of transitions \(\transR = 1\to 2\) and \(\transL = 2\to 1\). From the analytical equations derived in Section~\ref{sec:mainres} we show (b) conditional and unconditional probabilities of transitions, and the value of traffic rate, inter-transition time probability densities for (c) repeated and (d) alternated transitions. (e) In terms of a bias parameter, entropy production \(\sigma\) in a solid black curve, thermodynamic uncertainty relation's lower bound \(\sigma_\text{TUR}\) in a magenta solid curve, results from a Gillespie simulation \(\sigma_\mathcal{L}^\text{simu}\) are shown in {\color{red}blue crosses}; followed by the present results of inferred entropy production rate \(\sigma_\mathcal{L}\) in solid blue, and its decomposition in inter-transition times \(\sigma_t\) in dotted blue and sequence of transitions \(\sigma_\ell\) in dashed blue. The vertical line is the value of the (dimensionless) bias parameter for which the visible current vanishes. More details: transition rates are \(W_{24} = W_{34} = W_{41} = W_{43} = 1\), \(W_{14} = W_{21} = W_{32} = W_{42} = 20\) and \(W_{12}\) equals the exponential of the bias parameter; in (b)-(e) the bias parameter is fixed to 8.5; simulations were performed with a Gillespie algorithm for {\color{red}\(2\times 10^6\)s, the Kullback-Leibler divergence of inter-transition times was obtained with an unbiased estimation scheme [cf. main text].}}
    \label{fig:grid2}
\end{figure*}

The probability of a trajectory can be written in terms of the conditional probabilities of consecutive transitions and waiting times \(P ( t_i, \ell_i \vert \ell_{i-1})=P(t_i\vert \ell_{i-1},\ell_{i})P(\ell_i\vert \ell_{i-1})\), hence
\begin{equation}
    P[\Gamma_\tau^\mathcal{L}] = P(t_0, \ell_0) P(t_1, \ell_1\vert \ell_0) \cdots  P(t_n, \ell_n\vert \ell_{n-1}),
\end{equation}
\begin{equation}
    P[\overline{\Gamma}_\tau] = P(t_{n+1}, \overline{\ell}_n) P(t_{n}, \overline{\ell}_{n-1}\vert  \overline{\ell}_{n}) \cdots  P(t_1, \overline{\ell}_0\vert \overline{\ell}_{1}).
\end{equation}

After working out these expressions, see Appendix~\ref{infdetails} for details, we derive the following decomposition of the irreversibility measure \(\sigma_\mathcal{L}\):
\begin{equation}\label{epr_inference}
    \sigma_{\mathcal{L}} = \sigma_\ell + \sigma_t,
\end{equation}
where the first term is the contribution from sequence of transitions
{\color{red}
\begin{equation}\label{gen_epr_ell}
    \sigma_\ell = \traff \sum_{\ell, \ell' \in \mathcal{L}} P(\ell \vert \ell') P(\ell') \ln \frac{P(\ell\vert \ell')} {P(\overline{\ell'}\vert \overline{\ell})},
\end{equation}
}
and the second from the inter-transition times
{\color{red}
\begin{equation}\label{gen_epr_t}
    \sigma_t = \traff \sum_{\ell, \ell' \in \mathcal{L}} P(\ell \vert \ell') P(\ell') D\left[ P(t\vert \ell', \ell) \vert \vert P(t\vert \overline{\ell}, \overline{\ell'}) \right],
\end{equation}
where the indices in \(\sum_{\ell, \ell' \in \mathcal{L}}\) run over the set \(\mathcal{L}\) of all visible transitions.}

We now focus on the case of a system where two transitions in opposite directions between the same pair of states are visible \(\mathcal{L} = \{\transR, \transL\}\), as in single current monitoring. Notice that in this case the time-reversal of a transition is the also visible opposite transition \(\overline{\ell} = -\ell\). Thus above split of terms simplify to
\begin{align}\label{epr_alpha}
    \sigma_\ell = \traff [P(\transR) - P(\transL)] \ln \frac{P(\transR \vert \transR)}{P(\transL \vert \transL)} = J_\mathcal{L} A_\text{eff},
\end{align}
and
\begin{align}\label{epr_t}
    \sigma_t = & \traff P(\transR\vert\transR) P(\transR) D [ P(t\vert \transR,\transR) \vert \vert  P(t\vert \transL,\transL)]  \nonumber\\&+ \traff P(\transL\vert\transL)P(\transL) D [P(t\vert \transL,\transL) \vert \vert  P(t\vert \transR,\transR)].
\end{align}
The current over the observed transition is \(J_\mathcal{L} \coloneqq \traff [P(\transR) - P(\transL)]\) (cf. Appendix~\ref{infdetails}). In view of the usual bilinear form of the entropy production rate in usual nonequilibrium thermodynamics we identify the effective affinity \(A_\text{eff} \coloneqq \ln P(\transR \vert \transR ) / P(\transL \vert \transL )\).

One striking implication of Eq.~\eqref{epr_inference} is that the pairs of alternated transitions \(\transR\transL\) and \(\transL\transR\) do not play any role in the inferred entropy production. However, the incidence of repeated transitions \(\transR\transR\) and \(\transL\transL\) and their inter-transition times contribute to it. This means that only the statistics related to \(\transR\transR\) and \(\transL\transL\) are relevant to irreversibility.

{\color{red} The fact that both \(\sigma_\ell\) and \(\sigma_t\) are  linear combinations of Kullback-Leibler divergences with positive coefficients, implies that they are both always equal or greater than zero. This implies that both \(\sigma_\ell\geq 0\) and \(\sigma_t\geq 0\) are lower bounds to the rate of entropy production on their own. 
At equilibrium \(\sigma = \sigma_\ell=\sigma_t=0\), thus no irreversibility can be  detected from transition frequencies neither from inter-transition times. Out of equilibrium however \(\sigma_\ell\) and \(\sigma_t\) can vanish in different scenarios, which can be illustrated   for the case of observing a single pair of transitions: \(\sigma_\ell=0\)  when no net current (computed from frequency of transitions) is found along the visible transition  and \(\sigma_t=0\) when the Markov network is unicyclic, i.e. has a ring-like shape, as we show below. In addition, as proved in Ref.~\cite{seifertarxiv}, if the hidden network either has no cycles or satisfies detailed balance, one also gets  \(\sigma_t=0\).} 

Estimates of entropy production and irreversibility can be extracted from the statistics of single stationary trajectories. A recent example is the thermodynamic uncertainty relation, which allows estimating entropy production from empirical time-integrated currents without knowing the transition rates from the bound
\begin{equation}
    \sigma_\text{TUR} \coloneqq  \frac{2 \langle J\rangle^2}{ \mathrm{Var}(J)}\leq \sigma,
\end{equation}
which states that the entropy production rate is lower bounded by the average and variance of any stationary current \(J = \lim_{t\to\infty} \langle \sum_{i<j} d_{ij} n_{ij}(t)\rangle /t\) flowing over the system \cite{barato2015thermodynamic, gingrich2016dissipation}, with \(d_{ij}\) being the asymmetric current increment related to transition \(j\to i\) and \(n_{ij}(t)\) the number of such transitions in a time interval \(t\). 
For each trajectory, the stochastic time-integrated current \(J\) depends on the number of transitions in each direction, hence the full statistics of {\color{red}the} sequence of transitions should contain at least the same amount of information as the statistics of \(J\), therefore we conjecture \(\sigma_\text{TUR} \leq \sigma_\ell\). Furthermore, the intertransition times contribute to the entropy production rate and {\color{red}go} unnoticed by \(\langle J \rangle\) and \(\mathrm{Var}(J)\), therefore the contribution \(\sigma_t\) contains additional information such as the detection of irreversibility in the absence of net currents.

Fig. \ref{fig:grid2} illustrates how the entropy production inference is obtained using empirical estimates of \(P(\pm\vert \pm)\) and \(P(t\vert \pm,\pm)\) as a function of a bias parameter, a value present in transition rates that controls the {\color{red}preference for the performance of a counter-clockwise   cycle}. In a four-state multicyclic network, panel (e) shows the entropy production rate \(\sigma\) (solid black line) that is indeed larger than both \(\sigma_\ell\) and \(\sigma_t\). The contribution from {\color{red}the} sequence of transitions \(\sigma_\ell\) (dashed blue) coincides with the thermodynamic uncertainty relation \(\sigma_\text{TUR}\) (solid magenta) and they vanish for a value of bias parameter that stalls the current between \(1\leftrightarrow 2\), which is know as stalling force. The inter-transition time contribution is less sensitive to the bias parameter in this region. It does not vanish at the stalling force, leading to the detection of irreversibility when no net current is visible.

{\color{red}Lastly, for different values of bias parameter, a single trajectory of visible transitions and inter-transition times from Gillespie simulations was analyzed in view of Eqs.~\eqref{epr_alpha} and \eqref{epr_t} to obtain the inferred entropy production rate \(\sigma_\mathcal{L}^\text{simu}\) (blue crosses), in good agreement with the analytical \(\sigma_\mathcal{L}\). Notably, to tackle possible statistical biases that may arise in    \(\sigma_t\)  from crude histogram-counting procedures,  we rather employed the P\'erez-Cruz numerical method~\cite{4595271} that minimizes the statistical bias  in the estimation of Kullback-Leibler divergences. See~\cite{Harunari_KLD_estimation} for our open-source toolbox implementing our estimate of entropy production.   Further details of the implementation and convergence analyses are  discussed in Appendix~\ref{app:convergence}. }

\subsection{Ring networks}\label{eprunicyclic}

\begin{figure*}[ht]
    \centering
    \includegraphics[width=.95\textwidth]{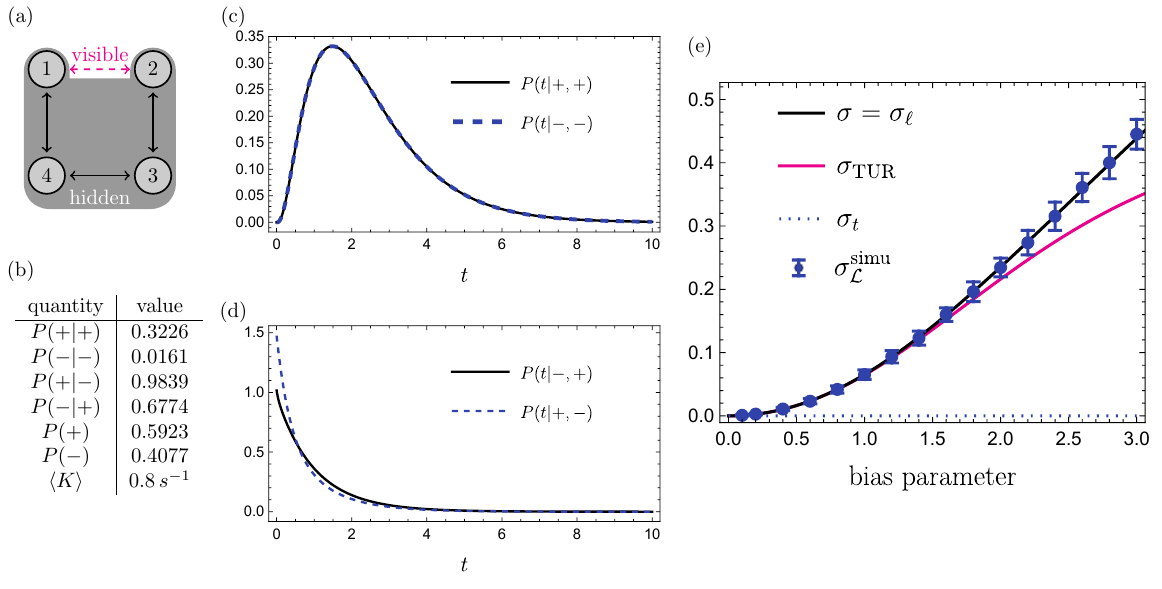}
    \vspace{-10pt}
    \caption{For a ring with four states, ring network, and visible transitions \(\transR = 1\to 2\) and \(\transL = 2\to 1\), as illustrated in (a), we show in (b) the conditional and unconditional probabilities of transitions and the visible traffic rate. (c) is the coinciding inter-transition time densities for repeated transitions and (d) for alternated. (e) is a summary of the entropy production rate inference scheme: entropy production \(\sigma\) and the sequence of transitions contribution \(\sigma_\ell\) coincide, both in solid black; inter-transition time contribution \(\sigma_t\) in dotted blue is shown to vanish; the thermodynamic uncertainty relation \(\sigma_\text{TUR}\) is depicted in solid magenta and simulations \(\sigma_\mathcal{L}^\text{simu}\) in blue dots with error bars. All transitions rates are equal to \(1\) apart from \(W_{14}\) that is the exponential of the dimensionless bias parameter, and entropy production rate dimensions are \(k_\text{B}/s\).}
    \label{fig:ring_inferences}
\end{figure*}

Networks with a ring topology are an important particular case for the inference of irreversibility. It has only one cycle and, therefore, one macroscopic flux and one affinity (thermodynamic force) \cite{RevModPhys.48.571}, such flux can be obtained from the solution of the master equation and the affinity is the logarithm of the product of all transition rates \(\ln \prod_i (W_{i+1,i}/ W_{i, i+1})\). We show that in this case the sequence of transitions contribution to the inferred entropy production in Eq.~ \eqref{epr_alpha} provides the exact real entropy production rate, ruling out the necessity of assessing all the microscopic details of stationary probabilities and transition rates.

In this case the stochastic matrix has a tridiagonal structure plus two terms on its corners \(W_{1N}\) and \(W_{N1}\), without loss of generality let us consider that \(1 \leftrightarrow 2\) is the observed transition, hence \([\mathbf{S}]_{ij} = W_{ij} (1-\delta_{ij,12})(1-\delta_{ij,21}) \). Due to the particular structure of \(\mathbf{S}\) in a ring, the Laplace expansion of its inverse leads to the fact that the effective affinity (associated with the visible transitions) equals in this case to the cycle affinity \(A\),
\begin{equation}
    A_\text{eff} = \ln \frac{ P(\transR\vert \transR) }{P(\transL\vert \transL)} = \ln \prod_i \frac{W_{i+1,i}}{W_{i,i+1}} = A,
\end{equation}
Analogously, we have found that this is also the case for the macroscopic affinity, which can be obtained from the ratio of conditional transition probabilities or estimated by their respective empirical frequencies.

In this case \(\sigma_\ell = J_\mathcal{L} A\), which is the definition of entropy production \(\sigma\) in a cycle. Adding this to the fact proven in Section \ref{sec:ring} that \(P(t\vert \transR,\transR) = P(t\vert \transL,\transL)\), implying \(\sigma_t =0\), we find that the inequality between inferred and real entropy production rate will be saturated and given solely and exactly by the sequence of transitions
\begin{equation}\label{unicyclicepr}
    \sigma_{\mathcal{L}} = \sigma_\ell = \sigma.
\end{equation}
In other words, the full entropy production of a ring network can be assessed by a single experiment in which a marginal observer collects statistics of {\color{red}the transitions between a single pair of states.}

Fig. \ref{fig:ring_inferences} shows the entropy production inference scheme for a ring network of four states. The contribution \(\sigma_t\) vanishes for any value of bias parameter due to the equality of inter-transition time densities for repeated transitions shown in panel (c). The values of \(\sigma\) and \(\sigma_\ell\) are precisely the same (solid black) as discussed in Eq.~\eqref{unicyclicepr}. Meanwhile, \(\sigma_\text{TUR}\) provides a lower bound that is approximately saturated for vanishing values of bias parameter, which represents the close to equilibrium regime.

\section{Inferences from visible transition in bio-molecular systems}\label{sec:bio}
Here we apply our theoretical framework to bio-molecular machines where partial information, stemming from the observation of {\color{red}a} few transitions, is experimentally accessible.  For example, DNA polymerase~\cite{10.1093/nar/gkv204}, data obtained from single-molecule FRET microscopy \cite{Verbrugge2007,Shi_FRET} and optical tweezers~\cite{Wen2008,Bustamante2021} to resolve the displacement of a motor along a track, yet most of the structural and chemical degrees of freedom are hidden. Inspired by these experimental limitations, we first focus on two examples of biologically-relevant molecular machines in which  we assume that one can only resolve mechanical transitions involving spatial displacements dynein (Sec.~\ref{sec:dynein}) and kinesin (Sec.~\ref{sec:kinesin}), which serve as case studies of ring and multicyclic networks, respectively.
Next, we  extend our study to motors that move in {\color{red}heterogeneous} tracks, and study the effect of the degree of disorder in the statistics of transitions (Sec.~\ref{sec:disorder}).

\subsection{Dynein ring model}
\label{sec:dynein}
Dyneins are cytoskeletal nano-scale motors that move along microtubules inside cells and perform a varied range of functions, like intracellular cargo transport and beating of flagella \cite{Canty2021, howard2001mechanics}. dyneins  transduce chemical energy from ATP hydrolysis into mechanical work done by displacing loads along the {\color{red}microtubule}. 

\begin{figure}[t!]
    \centering
    \includegraphics[width=\columnwidth]{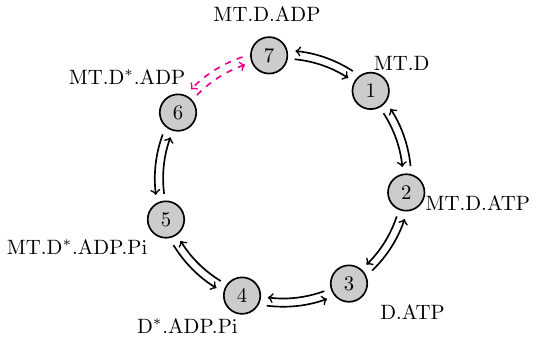}
    \caption{Sketch of the chemo-mechanical ring network for dynein with visible transitions \(6 \leftrightarrow 7\). The meaning of the transitions between each {\color{red}state} and the experimentally inferred values of the transition rates are listed in Table.~\ref{tab:dynein}.}
    \label{fig:dyn_cycle}
\end{figure}

Here, we study a unicyclic seven-state kinetic model of dynein stepping (cf. Fig.~\ref{fig:dyn_cycle}) that has a ring topology and is described in Refs.~\cite{vsarlah2014winch, hwang}. During every forward stepping cycle, one ATP molecule binds to the dynein (D) (1$\rightarrow$2), thereby triggering the release of the dynein from the microtubule (MT) (2$\rightarrow$3). This is followed by the hydrolysis of ATP that induces a conformational change of the dynein(D\(^*\)) (3$\rightarrow$4) and consequently leads to microtubule binding (4$\rightarrow$5). In the next step, release of one phosphate group Pi (5$\rightarrow$6) is followed by a power stroke (6$\rightarrow$7) and release of one adenosine diphosphate molecule ADP (7$\rightarrow$1). The different transition rates between these discrete states and their description are listed in Table~\ref{tab:dynein}.

We consider the setting where single molecule experiments can follow the cargo displacement and therefore observe only transitions \(6\leftrightarrow 7\). As discussed in Section~\ref{eprunicyclic} the inferred entropy production rate for this model is exactly given by \(\sigma=\sigma_\ell\) from Eq.~\eqref{epr_alpha}. From the network topology and transition rates we  evaluate \(\sigma_\ell\) analytically for different values of parameters such as the concentrations of ATP and ADP. 
The probabilities of a sequence of two transitions \(P(\pm\vert \pm)\) and \(P(\pm\vert \mp)\) are given from our framework by Eq.~\eqref{transprob}, and the probability of a single transition is given by Eq.~ \eqref{uniquetrans}.

\begin{figure}
    \centering
    \includegraphics[width=.45\textwidth]{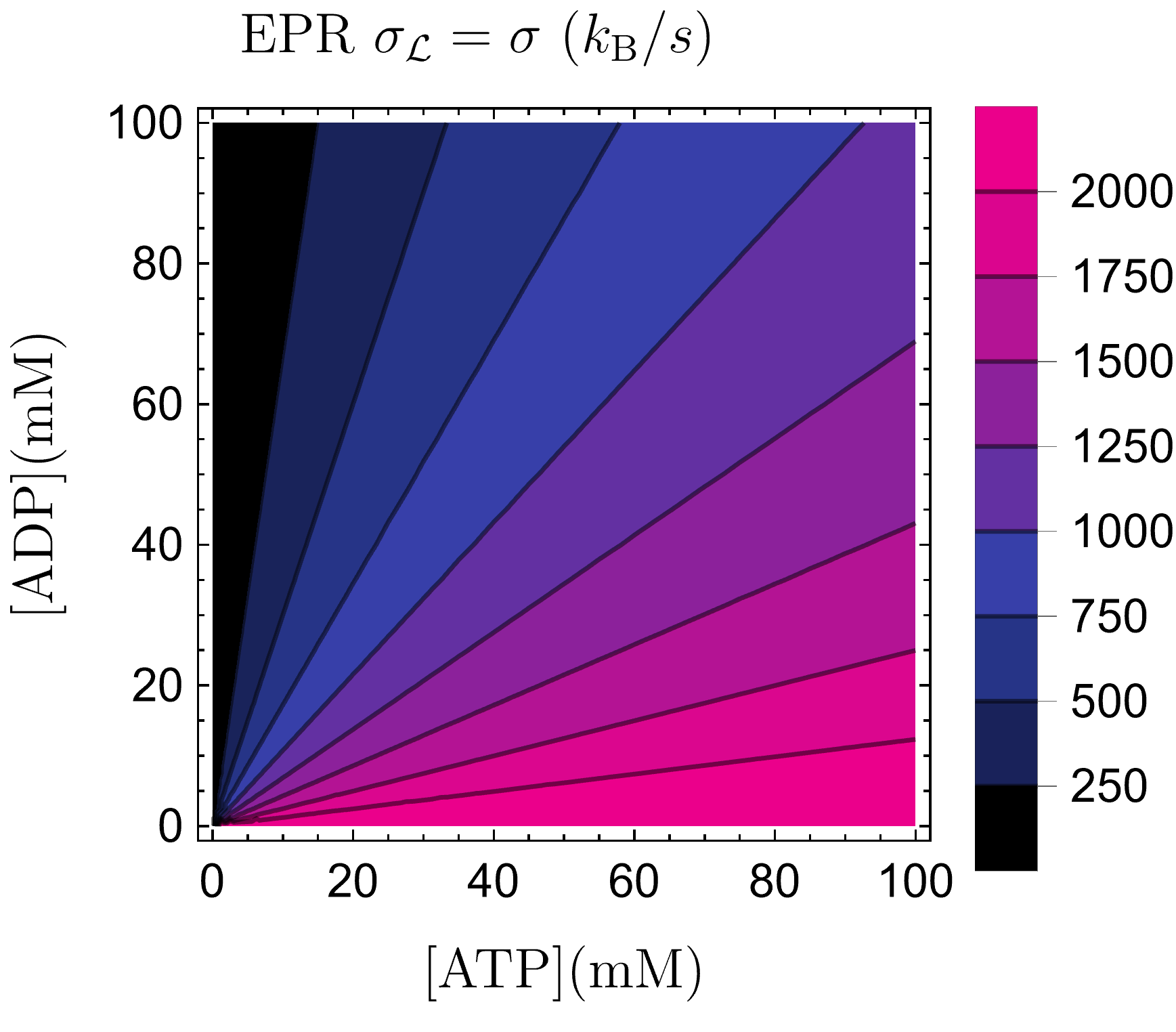}
    \caption{Entropy production rate in \(k_\text{B}/s\) for the dynein with visible transitions \(6 \leftrightarrow 7\), using rates from Table~\ref{tab:dynein} and [Pi]=1mM, in terms of [ATP] and [ADP].}
    \label{fig:dynein_ep}
\end{figure}

In Fig. \ref{fig:dynein_ep}, we observe that entropy production rate increases with the concentration of ATP and {\color{red}decreases} the concentration of ADP. This implies that the forward step of dynein is associated with high dissipation compared to the backward step. The typical dissipation rate for biophysical systems of nanometer to micrometer size ranges between 10-1000 $k_BT/s$ \cite{Bustamante2005}. Some examples are of kinesin  with dissipation rate 250 \(k_\text{B}T/s\) and single RNA hairpin with dissipation rate between 10-250 \(k_\text{B}T/s\). 

\begin{table}[htb]
    \centering
        \begin{tabular}{c c c}
            \hline Parameter & Description & Value  \\
            \hline
            
            {\color{red}\(W_{17}\)} & ADP release & 160 \\
            {\color{red}\(W_{71}\)} & ADP binding & 2.7$\times$[ADP] \\
            
            {\color{red}\(W_{21}\)} & ATP binding & 2$\times$[ATP] \\
            {\color{red}\(W_{12}\)} & ATP release & 50 \\
            
            {\color{red}\(W_{32}\)} & MT release in poststroke state & 500 \\
            {\color{red}\(W_{23}\)} & MT binding in poststroke state & 100 \\
            
            {\color{red}\(W_{43}\)} & linker swing to prestroke & 1000 \\
            {\color{red}\(W_{34}\)} & linker swing to poststroke & 100 \\
            
            {\color{red}\(W_{54}\)} & MT binding in prestroke state & 10000 \\
            {\color{red}\(W_{45}\)} & MT release in prestroke state & 500 \\
            
            {\color{red}\(W_{65}\)} & Pi release & 5000 \\
            {\color{red}\(W_{56}\)} & Pi binding & 0.01$\times$[Pi] \\
            
            {\color{red}\(W_{76}\)} & Power stroke & 5000 \\
            {\color{red}\(W_{67}\)} & Reverse stroke & 10 \\
            
            \hline
        \end{tabular}
	\caption{Transition rates for the chemo-mechanical cycle for the dynein model in Fig.~\ref{fig:dyn_cycle} (see~\cite{vsarlah2014winch, hwang}). All the rate constants $W_{i j}$ (except $W_{12}$, $W_{32}$ and $W_{67}$ which are in s$^{-1}\mu$M$^{-1}$)  are given in units of $s^{-1}$,  and the concentrations in  $\mu \mathrm{M}$. 
	Here MT refers to the microtubule.}
	\label{tab:dynein}
\end{table}

\subsection{Kinesin multicyclic model}
\label{sec:kinesin}

\begin{figure}[t!]
    \centering
    \includegraphics[width=.6\columnwidth]{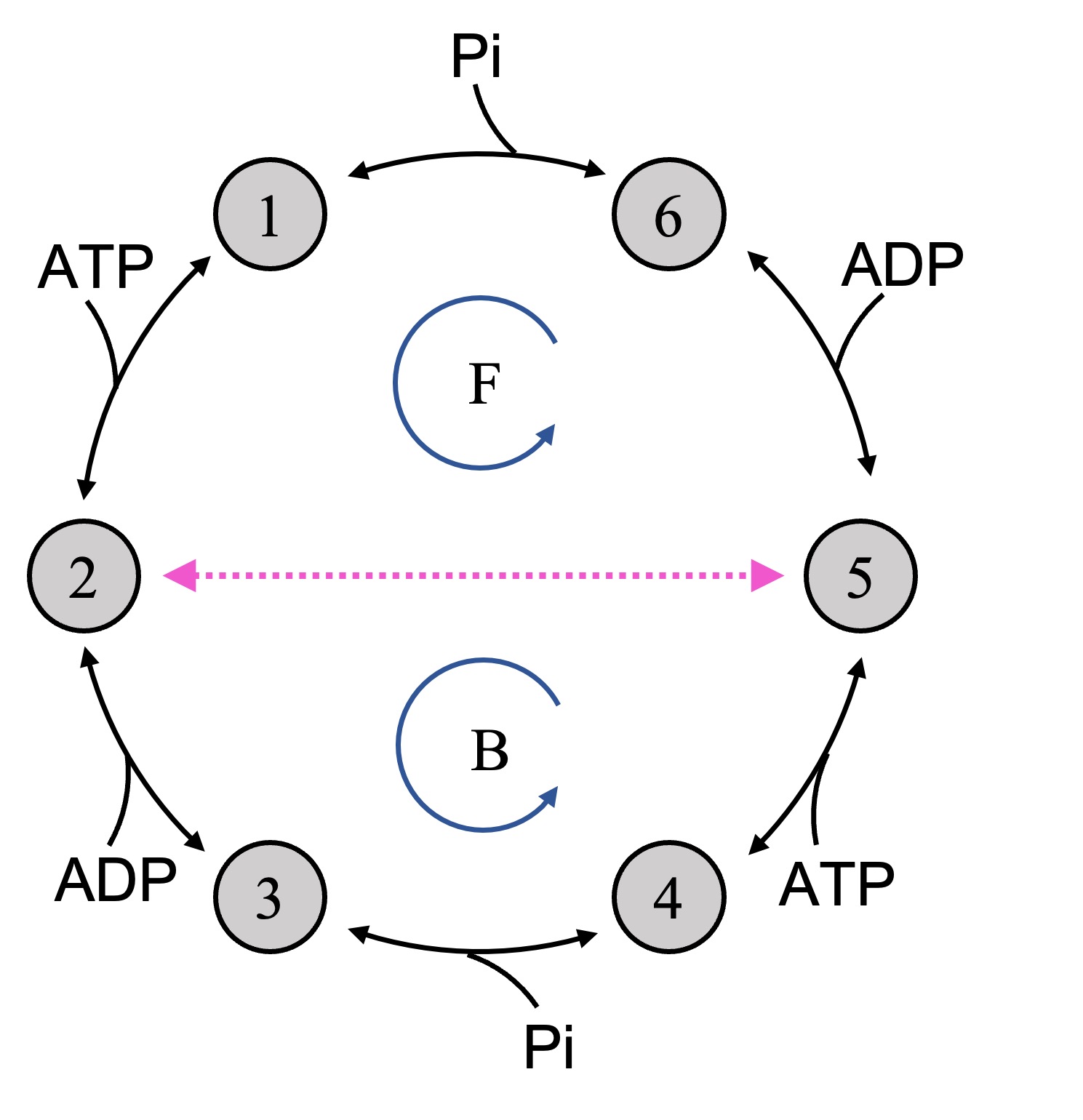}
    \caption{Sketch of the  chemo-mechanical network model used to describe  kinesin motion. The only visible transitions \(2\leftrightarrow 5\) are marked in dotted magenta. Here \(F\) and \(B\) denote the cycle corresponding to the forward and backward movement of kinesin, respectively.}
    \label{fig:kinesin_network}
\end{figure}

We now study a stochastic model  for  kinesin  motion~\cite{PhysRevLett.98.258102} validated in single-molecule experimental studies~\cite{carter2005mechanics, nishiyama2002chemomechanical}, see  Fig.~\ref{fig:kinesin_network} for an illustration. The model is described by a chemo-mechanical network comprising six discrete states which describe the mechanism of movement of kinesin on the microtubule. Notice that it has two independent cycles: ``F'' cycle $[(1) \rightarrow$(2) $\rightarrow(5) \rightarrow(6)\rightarrow(1)]$  corresponding to the forward  motion of kinesin by one step, and ``B'' cycle $[(4) \rightarrow(5) F\rightarrow(2) \rightarrow(3) \rightarrow(4)]$  resulting in a step backwards.  The dynamics along one F cycle is as follows: after ATP binding ($1\to 2$), kinesin makes a step forward ($2\to 5$) in the filament, followed by ATP hydrolysis that results in the release of one ADP molecule ($5\to 6$) and inorganic phosphate Pi ($6\to 1$). The backward B cycle proceeds similarly, with the only difference that after the binding of ATP to kinesin a backward step along the filament ($5\to 2$) occurs.
Notice that, in contrast to the model example of dynein, here forward and backward movements are driven by the hydrolysis of one molecule of ATP. The transition rate values are listed in the Table~\ref{tab:kinesin}. An external load force $f$ biases the transition rates $W_{25}$ and $W_{52}$ involving spatial motion:
    \begin{eqnarray}
        W_{52}(f)&=&W^0_{52} \mathrm{e}^{-\theta f d_{0} / k_{\mathrm{B}} T}\nonumber\\\relax
        W_{25}(f)&=&W^0_{25} \mathrm{e}^{(1-\theta) f d_{0} / k_{\mathrm{B}} T},
    \end{eqnarray}
where $\theta$ is the load distribution factor, $d_0$ is the step size and $f$ is the load force. On the other hand, for the chemical transitions we have 
    \begin{equation}
        W_{i j}(f)=2 W^0_{i j}(1+e^{\chi_{i j} f d_{0} / k_{\mathrm{B}} T})^{-1},
    \end{equation}
where $\chi_{i j}$ represents the mechanical strain on catalytic domains with $\chi_{i j}=\chi_{j i}>0$ where $i, j \neq 2,5$ and {\color{red}the concentration of molecular species involved in the chemical transitions are accounted  in most of the  \(W_{ij}^0\) rates, see Table\ref{tab:kinesin}}.
  
We now focus on the statistics of  the transitions associated with the mechanical movement of kinesin i.e. \(2\leftrightarrow 5\), which are the only ones that can be observed experimentally. For our calculations, we have considered the concentration for ADP to be  [ADP]$=70 \mu$M and $[{\rm P}_i]=$1mM, the load distribution factor $\theta=0.65$, $d_0=2 k_{\mathrm{B}}T$, $\chi_{12}=0.25$ and $\chi_{56}=\chi_{61}=0.15$.

\begin{table}[htb]
    \centering
        \begin{tabular}{c c c}
            \hline Parameter & Description & Value  \\
            \hline
            
            \(W^0_{21} = W^0_{54}\) & ATP binding & 2.0$\times$[ATP] \\
            \(W^0_{12}\) & Release of ATP & 100 \\
            
            \(W^0_{32} = W^0_{65}\) & ADP release & 100 \\
            \(W^0_{23} = W^0_{56}\) & ADP binding & 0.02$\times$[ADP] \\
            
            \(W^0_{25}\) & ATP binding & 0.24 \\
            \(W^0_{52}\) & Mechanical step & $3\times10^{5}$\\
            
            \(W^0_{43} = W^0_{16}\) & Hydrolysis of ATP & 100 \\
            \(W^0_{34} = W^0_{61}\) & Pi binding & 0.02$\times$[Pi] \\
            
            \(W^0_{45}\) & Release of ATP & ${\color{red}W}^0_{12} ({\color{red}W}^0_{25}/{\color{red}W}^0_{52})^2$ \\

            \hline
        \label{tab:kinesin}
        \end{tabular}
	\caption{Transition rates for the chemo-mechanical cycle for kinesin model in Fig. \ref{fig:kinesin_network} \cite{PhysRevLett.98.258102}. All the rate constants $W_{i j}$ (except $W_{21}$, $W_{23}$, $W_{34}$, $W_{54}$, $W_{56}$ and $W_{61}$, which are in s$^{-1}\mu$M$^{-1}$) are given in units of $s^{-1}$, and the concentrations in $\mu \mathrm{M}$.}
\end{table}

\begin{figure}[t!]
    \centering
    \includegraphics[width=.5\textwidth]{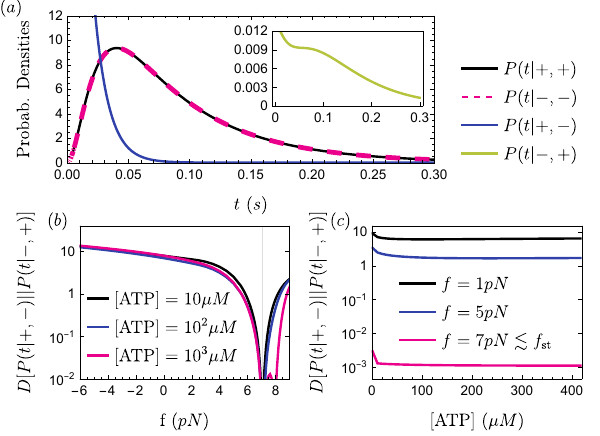}
    \caption{Exact inter-transition time statistics for the kinesin model in Fig. \ref{fig:kinesin_network} with visible transitions \(2 \leftrightarrow 5\). (a) Inter-transition time densities for every possible pair of transitions. (b) Kullback-Leibler divergence for alternated transitions in terms of the external force \(f\) with vertical line corresponding to the stalling force of \(f_\text{st}\sim 7.02\ pN\), (c) and in terms of the ATP concentration. The rates are displayed Table \ref{tab:kinesin} and \(\text{[ADP]} = \text{[P]} = 5 \mu M\).}
    \label{fig:kinesin_wtd}
\end{figure}

Fig. \ref{fig:kinesin_wtd} shows the inter-transition statistics of this model obtained from the analytical expressions in Section~\ref{sec:trans_stat}, which displays a rich structure due to the multicyclic structure of the model. {\color{red}Our results show that, apart from being defined in a network with two cycles, inter-transition time densities for repeated transitions are identical \(P (t\vert \transR, \transR) = P (t \vert \transL, \transL)\) [Fig. \ref{fig:kinesin_wtd}(a)]. This property results from the symmetry property that the F and B cycles of the model pass through transitions with identical rates; it is not a generic property for multicyclic networks (see Fig.~\ref{fig:grid2}c). As can be seen in Fig. \ref{fig:kinesin_wtd}(a), the inter-transition times have very different densities, which is due to the transition rates being orders of magnitude apart. In this case, alternated transitions are much faster than repeated ones.}

As can be seen in Fig. \ref{fig:kinesin_wtd} (b), alternated transitions in general have different inter-transition time densities but, for the stall force, \(P(t\vert \transR, \transL)\) and \(P(t\vert \transL, \transR)\) become similar, as can be seen by minima in \(D [P(t\vert \transR, \transL) \vert \vert P(t\vert \transL, \transR)]\). Both for the force and the concentrations of ATP [cf. Fig. \ref{fig:kinesin_wtd} (c)] we observe regions of decreasing divergence, however the entropy production rate is increasing in these regions, this is an evidence of the finding that inter-transition times between alternated transitions do not contribute to the dissipation.

\begin{figure}[t!]
    \centering
    \includegraphics[width=.48\textwidth]{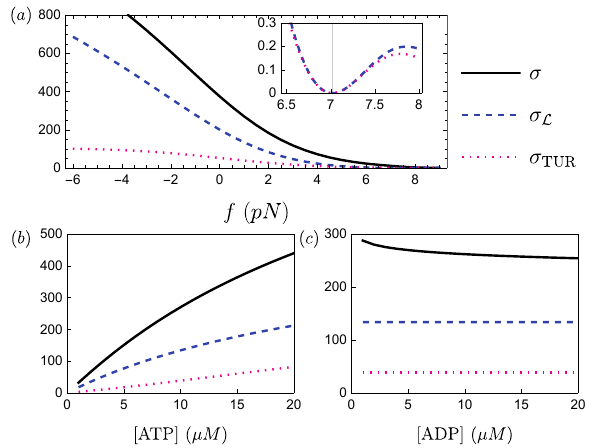}
    \caption{Exact analytical values for the rate of entropy production for the kinesin model in Fig. \ref{fig:kinesin_network} with visible transitions \(2 \leftrightarrow 5\):  entropy production \(\sigma\) of the underlying Markov chain (black curves), inferred entropy production from transition statistics \(\sigma_\mathcal{L}\) (blue dashed line) and  estimate from the  thermodynamic uncertainty relation lower bound \(\sigma_\text{TUR}\) (magenta dotted line). (a) Values in terms of the external load force \(f\) and a zoomed-in view around the stalling force in the inset, for this case [ATP]\(=10 \mu M\), and [ADP]=[P]\(=5 \mu M\). (b) In terms of the concentration of ATP, with \(f=1 pN\), and [ADP]=[P]\(=5 \mu M\). (c) In terms of the concentration of ADP, with \(f=1 pN\), [ATP]\(=10 \mu M\), and [P]\(=5 \mu M\).}
    \label{fig:kinesin_epr}
\end{figure}

Fig.~\ref{fig:kinesin_epr} shows entropy production rate \(\sigma\) and the values inferred from our approach of observing the forward and backward mechanical transition and the thermodynamic uncertainty relation. Fig.~\ref{fig:kinesin_epr}(a) is in terms of the external force \(f\) with a zoomed-in view around the stalling force, for which both \(\sigma_{\mathcal{L}}\) and \(\sigma_{\text{TUR}}\) vanish since there is no flux and no inter-transition time asymmetry between repeated transitions. Fig.~\ref{fig:kinesin_epr}(b) is depicted in terms of the concentration of ATP and (c) of ADP, from them we observe a monotonic increase of dissipation with [ATP] while it is almost independent of [ADP]. In this model \(\sigma_{\mathcal{L}}\) obtained from Eq.~\ref{epr_inference} in general provides a good estimate for \(\sigma\), in general overperforming the thermodynamic uncertainty relation \(\sigma_{\text{TUR}}\). Due to the absence of \(\sigma_t\) no dissipation is detected when no net current is present (at stalling force).

\begin{figure}[t!]
    \centering
    \includegraphics[width=\columnwidth]{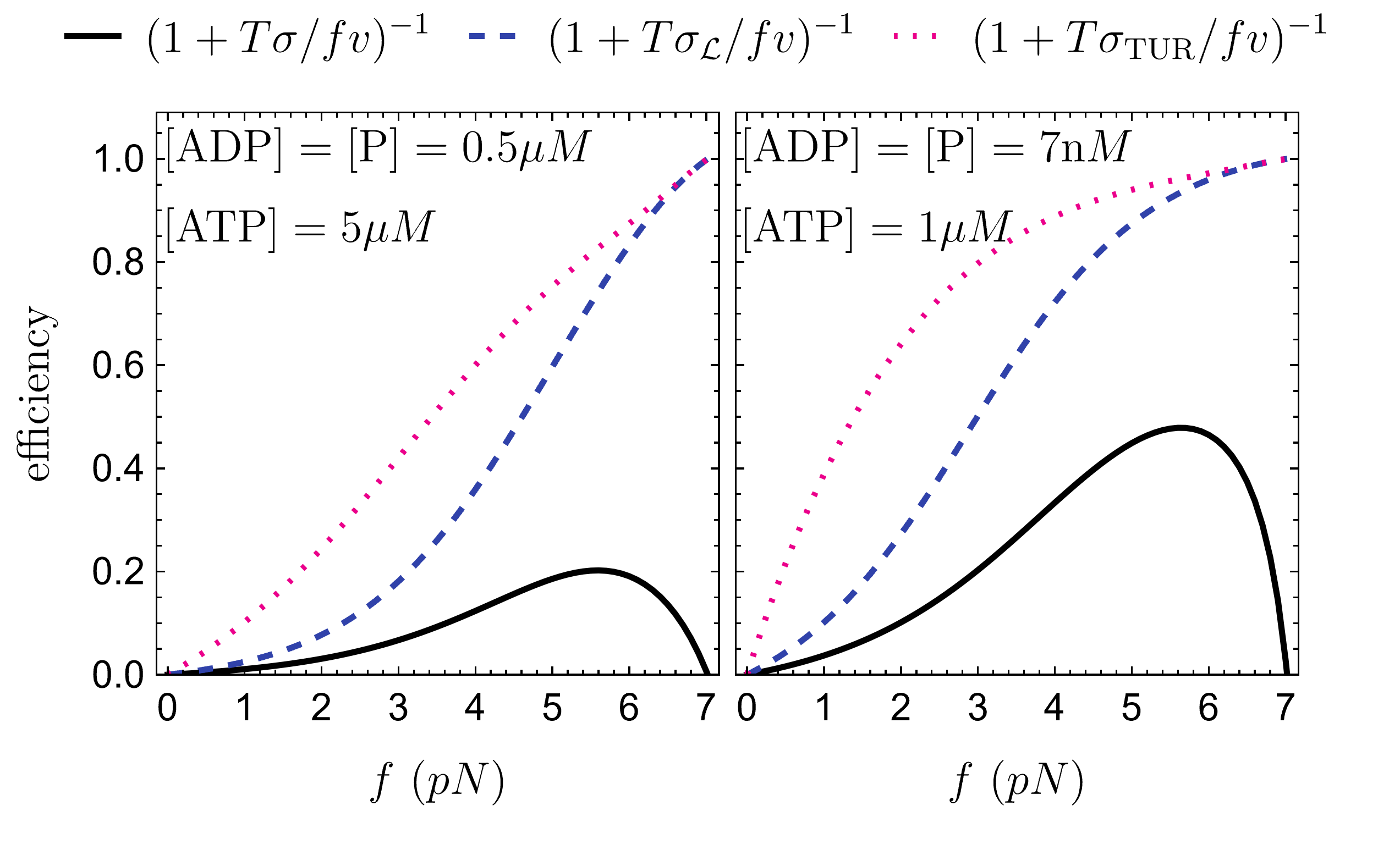}
    \caption{{\color{red}Efficiency (solid black) of a kinesin motor doing work against a force \(f\), and its upper bounds obtained from the visible entropy production (dashed blue) and the thermodynamic uncertainty relation (dotted magenta). More details in Section~\ref{sec:kinesin}.}}
    \label{fig:kin_eff}
\end{figure}

{\color{red}\subsection{Bounds for efficiency of molecular motors}

To date, one of the most remarkable applications of the thermodynamic uncertainty relation is the upper bounding of biological motors by the first and second moments of its motion \cite{Pietzonka_2016, Dechant_2018}. Here, we consider the specific class of  molecular motors in which the ``stepping transition'' does not involve chemical fuel consumption but work done against {\color{red}an} external load force, which include as specific examples the dynein and kinesin models of the previous sections.  Within this class of molecular motors, the rate of entropy production can be written as \(\sigma = (\dot{w}_\text{chem} - fv)/T\), where  \(\dot{w}_\text{chem}\) is the average power done on the motor by the chemical transitions (e.g. by  the ATP hydrolysis cycle). On the other hand,  \(fv\) is the average power exerted by the load force $f$, with \(v\) being the net velocity of the motor along the track. For such motors, the second law \(\sigma \geq 0\) implies that one can introduce a notion of  efficiency  as \(\eta = fv/\dot{w}_\text{chem}\), which can be expressed in terms of the entropy production rate as follows
\begin{equation}
    \eta = \frac{1}{1+T\sigma/fv}.
    \label{eq:effsigma}
\end{equation}
From our observations, we conjecture that the hierarchy of bounds $\sigma \geq \sigma_{\mathcal{L}}\geq \sigma_{\rm TUR}$, which implies together with Eq.~\eqref{eq:effsigma} the following conjectured hierarchy of upper bounds for the molecular motors' efficiency
\begin{align}\label{eq:eff_hierarchy}
    \eta \leq \frac{1}{1+T\sigma_\mathcal{L} /fv}\leq \frac{1}{1+T\sigma_{\rm TUR} /fv} \leq 1.
\end{align}
Equation~\eqref{eq:eff_hierarchy} implies that the present inference scheme leads to a tighter upper bound to the efficiency than that based on the thermodynamic uncertainty relation introduced in Ref.~\cite{Pietzonka_2016}. Unlike \(\sigma_\ell\) and \(\sigma_\text{TUR}\), \(\sigma_\mathcal{L}\) includes information about irreversibility through inter-transition times, which shows how the notion of time tightens the efficiency bound.

We illustrate the bounds~\eqref{eq:eff_hierarchy} in  Fig.~\ref{fig:kin_eff} for our model of kinesin.  Its efficiency is positive in the regime where load force and net movement have opposite signs \(0\leq f \leq f_\text{st}\) (\(f_\text{st} \approx 7.02\) is the stalling force), thus the motor performs work against the applied force at the cost of ATP consumption. For the parameter choices that we explored,  we observe that close to the motor maximum efficiency the upper bound obtained from transition statistics is a $13\%$ closer to the actual value with respect to the estimate obtained from the TUR.} 

\subsection{Motion on disordered tracks}
\label{sec:disorder}

In many instances, the stochastic motion of molecular machines display a disordered nature due to the heterogeneity of the track. For example, template-copying machines like DNA and RNA polymerases \cite{PhysRevLett.117.238101} and ribosomes \cite{Rudorf2015} are often modelled as machines whose motion is dependent on the sequence constituting the track, {\color{red}in such a way that the transition rates depend on the specific monomer type that the machine encounters at every step~\cite{PhysRevE.71.041906,PhysRevLett.79.2895}}. In this section, we study the effects of {\color{red} the track's disorder in the inter-transition statistics associated with the motion of a minimal stochastic model of a molecular machine.} 

\begin{figure}[t!]
	\centering
	\includegraphics[width=\columnwidth]{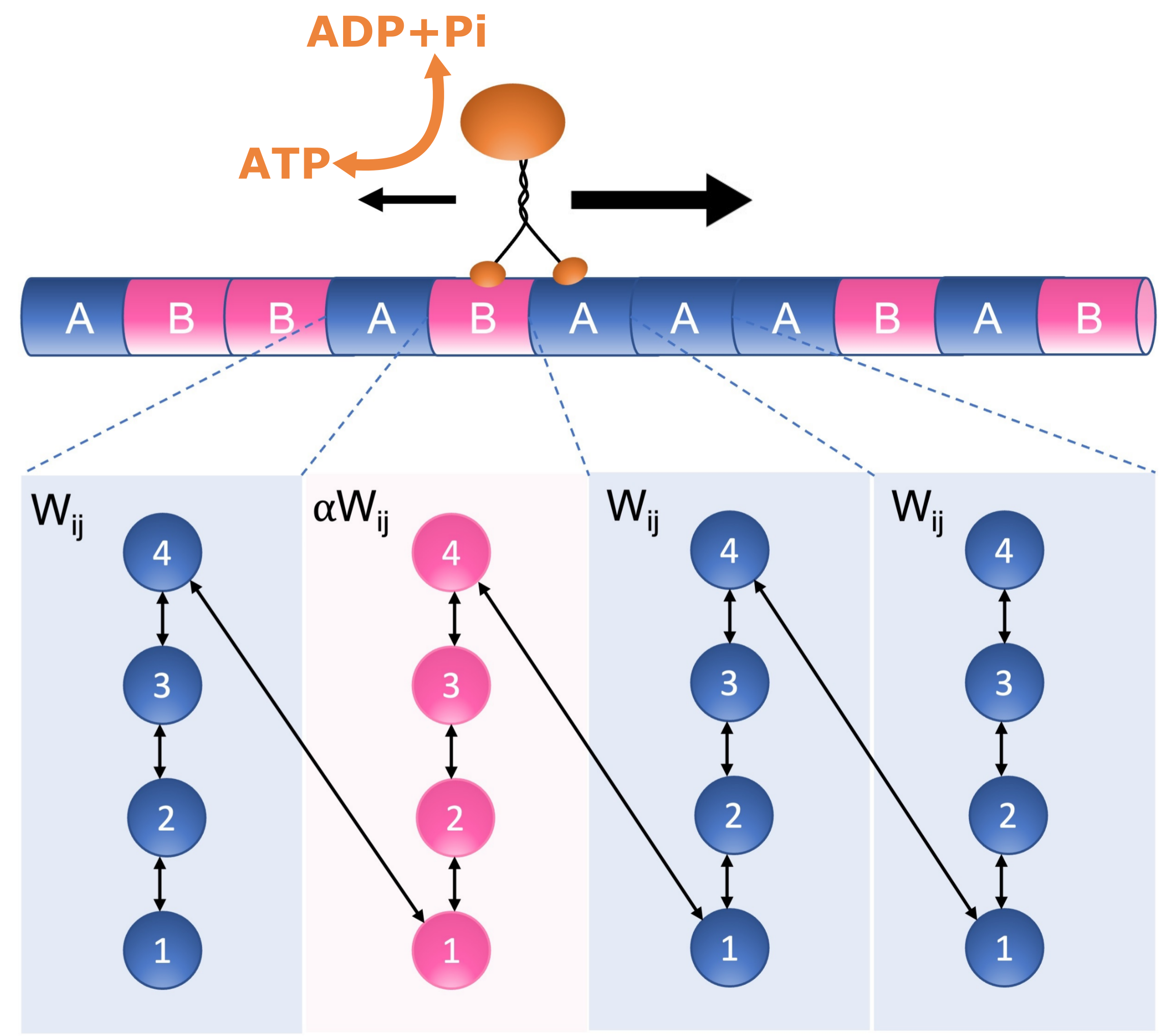}
	\caption{{\color{red}Sketch  of our  minimal stochastic model for molecular motor motion  in a disordered linear track composed of two types of monomer A and B. The track is generated as an i.i.d. sequence of monomers  \(A\) and \(B\) that occur  with probabilities \(p\) and \(1-p\) respectively.
	Within each periodicity cell of type \(A\) or \(B\), the motor internal states follow have the same structure given by a unicyclic network. The transition rates depend on the motor's position on the track according to Eq.~\ref{disorderrates}, where \(\alpha\) is a disorder factor. We also assume that the only visible transitions are those related to translocation to the right (step ``forward'') \(\transR = 4\to 1\) and to the left  (step ``backward'')  \(\transL = 1\to 4\).}}
	\label{fig:random_schematic}
\end{figure}

{\color{red} We consider a minimal stochastic model of a molecular machine that moves along a track by burning fuel (i.e. by hydrolysis of ATP). The machine undergoes a series of conformational changes and translocates  on a linear heterogeneous  track (a polymer) composed of two types of monomers, labeled \(A\) and \(B\). We assume that the track is infinite (i.e. we effectively have {\bf annealed} disorder), and that the generation of the template \(q_n \in \{A, B\}\), \(n=\{1,2,\ldots\}\), is an i.i.d. process such with prescribed probabilities $P\left(q_{n}=A\right)=p$ and $P\left(q_{n}=B\right)=1-p$ for the occurrence of A and B type monomers, respectively. For our numerical study, we generate the template before running the simulations and use the same template for every run. Figure~\ref{fig:random_schematic} sketches the disordered nature of a track along the motion of the molecular machine. We also assume that the motor moves following a unicyclic enzymatic reaction composed of four internal configurational states and that only two transitions are visible  \(4\to1=+\) and \(1\to4=-\) corresponding to forward and backward steps along the track, respectively. 
The template disorder is implemented in the stochastic model as follows: When the motor reaches a monomer of type~\(q_{n}\), its internal configurational states within one periodicity cell are connected by rates
\begin{equation}\label{disorderrates}
    W_{ij}^{(q_{n})} = \begin{cases} W_{ij} & \text{if }q_{n}=A\text{ and }i,j\neq 4,1 \\ \alpha W_{ij} &\text{if }q_{n}=B\text{ and }i,j\neq 4,1 \\ W_{41} & \text{if }q_{n-1}=A\text{ and }i,j=4,1 \\ \alpha W_{41} &\text{if }q_{n-1}=B\text{ and }i,j=4,1 \end{cases}
\end{equation}
where $\alpha\in (0,1]$ is the disorder factor. This factor scales the transition rates, effectively slowing or accelerating the transitions depending on the track position. As a convention, we have set the transition rate related to a back step \(W_{41}^{(q_{n})}\) to be defined in terms of the previous monomer's type \(q_{n-1}\). Apart from specific choices of the parameters, this motor has a nonequilibrium dynamics, evidenced by a net drift along the track. In ring topologies, we have observed that inter-transition times do not contain irreversibility traces, which is not necessarily true for the disordered case.}

\begin{figure}[t!]
	\centering
	\includegraphics[width=\columnwidth]{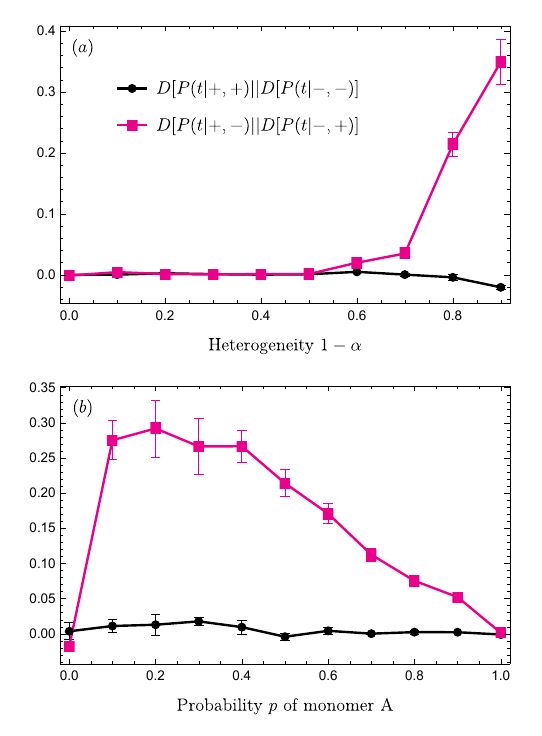}
	\caption{{\color{red}Kullback-Leibler divergence of the inter-transition time densities for repeated ($D[P(t\vert +,+)\vert\vert P(t\vert -,-)]$, black circles) and alternated ($D[P(t\vert +,-)\vert\vert P(t\vert -,+)]$, magenta squares) transitions obtained from numerical simulations of the model sketched in Fig.~\ref{fig:random_schematic}.
	We plot the values of these Kullback-Leibler divergences as a function of two parameters of disorder: (a) in terms of heterogeneity \(1-\alpha\) with fixed monomer probability $p=0.5$, and  (b) in terms of probability \(p\) of monomer A, with fixed \(\alpha = 0.2\). The rates used for Gillespie simulations are $W_{12} = W_{43}=1 \, s^{-1}$, $W_{21} = W_{23} = W_{34} =5 \, s^{-1}$ and $W_{32} = W_{14} = W_{41} =4 \, s^{-1}$, which leads to a nonequilibrium dynamics, and \(\alpha\) is introduced according to Eq.~\ref{disorderrates}. Error bars represent the standard deviation from five trajectories each of duration \(5\times 10^6\)s.}}
	\label{fig:KLDivergence}
\end{figure}

{\color{red}We now study how the disorder parameters \(\alpha\) and \(p\) of this minimal model affects the inter-transition time statistics of successive repeated transitions \(\transR\transR\) and \(\transR\transL\), and alternated transitions \(\transL\transR\) and \(\transL\transL\). From the simulation of a molecular motor on a disordered track with four internal states, we observe in Fig.~\ref{fig:KLDivergence} that only the statistics of  inter-transition times between alternated transitions are affected by the degree of disorder.  In particular,   we observe for our example model that   \(D[P(t \vert \transR, \transR) \vert \vert P(t\vert \transL, \transL)] = 0\) i.e. a symmetry relation between the inter-transition-time distributions of repeated transitions $P(t \vert \transR, \transR)\simeq P(t \vert \transL, \transL)$ which implies $\sigma_t\simeq 0$. As we saw in Sec.~\ref{eprunicyclic}, such symmetry relation is a hallmark of unicyclic networks whereas here we have effectively a multicyclic network with two types of cycles A and B. We expect the symmetry $P(t \vert \transR, \transR)\simeq P(t \vert \transL, \transL)$, which is already expected for the homogeneous case (\(\alpha = 1\), \(p=0\) or \(p=1\)),  to be  originated by the fact that different monomer types are just affecting the timescale of the jumps and not the internal network topology within each periodicity cell.

The results for repeated transitions are in stark contrast with our observations for the alternated transitions, see magenta curves in  Fig.~\ref{fig:KLDivergence}, which shows that  \(D[P(t\vert \transR, \transL) \vert \vert P(t\vert \transL, \transR)] \) is strongly dependent on the values of \(\alpha\) and \(p\) controlling the amount of disorder in the track.  Figure~\ref{fig:KLDivergence}a shows that the degree of asymmetry \(D[P(t\vert \transR, \transL) \vert \vert P(t\vert \transL, \transR)] \)  in the alternated inter-transition-time statistics increases monotonously with the degree of heterogeneity  affecting the timescale of the jumps given by $1-\alpha$.  On the other hand, \(D[P(t\vert \transR, \transL) \vert \vert P(t\vert \transL, \transR)]>0 \) is also able to probe the presence of sequence heterogeneity as we vary the probability of A monomers $p$ for a fixed $\alpha $, see  Fig.~\ref{fig:KLDivergence}b.  Note that in the latter case, we recover \(D[P(t\vert \transR, \transL) \vert \vert P(t\vert \transL, \transR)]\simeq 0 \) for the limiting cases $p=0$ and $p=1$, which correspond to homogeneous, unicyclic networks.  Taken together, these results highlight the possibility of using the inter-transition-time statistics between alternated transitions as a probe of the presence of underlying disorder in cyclic enzymatic reactions, which could be further generalized in future work.}



%
\section{Discussion}\label{sec:discussion}
%

{\color{red}In this work we  have developed  results for generic stationary Markov-jump processes, in and out of equilibrium, whose partial information is restricted to the observation of a partial set among all its network of transitions. In particular, we investigated the question: what can one learn from counting the frequency of a partial set of visible transitions and the time elapse between two such visible and successive transitions occurring in a time-series?  }
We have tackled the problem of learning dynamic and thermodynamic properties of a system in which only a few transitions are visible to the observer, a novel coarse-graining scheme that proves to be physically meaningful and that provides information through simple relations. For the broad class of stationary Markov processes, we have derived exact analytical results for the conditional and unconditional probability of occurrence of successive transitions and for the time elapsed between successive transitions (inter-transition times), which together comprise all the information available to an observer that can only track {\color{red}the} occurrence of {\color{red}a} few visible transitions. 

{\color{red} A key insight of our work is that measuring inter-transition times is crucial for thermodynamic inference.} Inter-transition time statistics  of two successive repeated transitions (e.g. + followed by +) {\color{red}carry} different information than that of two successive different transitions (e.g. + followed by -). Repeated {\color{red}transition} frequencies and inter-transition times contain information about time irreversibility, which can be used to establish tight lower bounds for entropy production even in the absence of probability currents in the transition state-space. Counter-intuitively, alternated transitions do not contribute to entropy production estimates, but their statistics provide means to identify the presence of disorder in the hidden state space. Taken together, our work unveils the relevance of inter-transition times in thermodynamic inference, putting forward recent works~\cite{martinez2019inferring, skinner2021estimating, doi:10.1073/pnas.0804641105,seifertarxiv} that identified footprints of irreversibility in asymmetries of waiting-time distributions in states rather than in transitions. 
{\color{red}Exploring symmetry properties and developing inference methods from statistics of a variety of waiting times is a promising novel area of research within the field of stochastic thermodynamics~\cite{neri2019integral, martinez2019inferring, skinner2021estimating, hartich2021comment, PhysRevX.11.041047, seifertarxiv}. In particular, the coetaneous manuscript~\cite{seifertarxiv} also reports an analysis of waiting-time statistics between transitions, and provides complementary results to those developed in our framework and applications.


The results we have obtained are generic and can be applied to large and complex networks, for any given set of visible transitions. One must notice that the inferences become limited when observing a tiny fraction of the transitions on very large networks. Therefore, it would be interesting to study how robust inferences are in relation to the visible portion of the network, in particular with large and complex structures. In the context of biological systems,  this hurdle may be overcome with  recent experimental developments. For example, using two-colour single-molecule photoinduced electron transfer fluorescence imaging microscopy \cite{Schubert2021} and three-colour FRET \cite{Yoo2018}, one can simultaneously probe multiple conformational changes within an individual bio-molecule using one fluorescence colour per coordinate. {\color{red}Additionally, the bias in the estimation of relative entropy is circumvented using an unbiased estimator~\cite{4595271}, whose implementation we made available  as an open-source code in Ref.~\cite{Harunari_KLD_estimation}. This open-source toolbox can be used to estimate Kullback-Leibler divergences from experimental time-series and, consequently, also the visible entropy production developed herein.}


A possible application of the present formalism is to the problem of making insightful considerations about the efficiency of complex biochemical systems or, more in general, of multiterminal systems with more than one input/output \cite{brandner2013multi}, or with unknown losses \cite{vroylandt2016efficiency}. In fact, while efficiency is well defined when there is one definite input and output, 
biochemical systems most often involve many sources. For example, in glycolysis one has ATP, ADP, lactate, water, phosphate, and glucose as metabolites \cite{rawls2019simplified}; being the universal energy tokens, it makes sense to consider the ratio of ADP to ATP production as a measure of efficiency, but then the problem is how to single them out of all other mechanisms and make claims about the efficiency of the process. Furthermore, in more complex biochemical networks, such as those that also involve respiration, one might also want to focus on other metabolites (oxygen, carbon dioxide etc.). To develop such an approach, it is thus mandatory to develop a more phenomenological theory that is consistent with the fundamental tenets of thermodynamics, but can also be adapted to the specific tasks/instruments that the observer has in mind. In this respect, the theory presented here may provide a general conceptual and operational scheme.}

We illustrated our results in two models of motion of molecular motors that have been validated with experimental data, revealing that our methodology could be applied to real data extracted from e.g. single-molecule experiments. We expect that our generic inference techniques will be applied to other disciplines where partially observed transitions emerge, such as diagnosis algorithms~\cite{sampath1996failure}, finite automata~\cite{wang2007algorithm,wolpert2019stochastic}, Markov decision processes~\cite{lovejoy1991survey}, disease spreading~\cite{bhadra2011malaria}, information machines \cite{serreli2007molecular}, probing of open (quantum) systems \cite{viisanen2015incomplete, PhysRevE.91.012145}, and Maxwell demons \cite{PhysRevLett.110.040601}.

%
\section*{Acknowledgments}
%

We are thankful to Ken Sekimoto and Nahuel Freitas for useful discussions.  We also thank Fahad Kamulegeya for preliminary numerical results. PEH acknowledges grants \#2017/24567-0 and \#2020/03708-8, S\~ao Paulo Research Foundation (FAPESP) and Massimiliano Esposito for the hosting in his group. PEH and AD acknowledge the financial support from the ICTP Quantitative Life Sciences section. MP acknowledges the National Research Fund Luxembourg (project CORE ThermoComp C17/MS/11696700) and the European Research Council, project NanoThermo (ERC-2015-CoG Agreement No. 681456).

%
\appendix
%

\section{Proof of Eq.~\eqref{theorem}}\label{proof}

We now prove Eq.~\eqref{theorem} in the Main Text as follows. We map the {\color{red}``first-transition time''} problem into a  first-passage time {\color{red}problem} by introducing auxiliary absorbing states for each transition in \(\mathcal{L}\). This procedure is inspired by recent work on first-passage times between states in a Markov chain~\cite{sekimoto2021derivation}, and  is also similar to the manipulation of networks to obtain current statistics by creating copies of some states introduced by Hill~\cite{hill1988interrelations, sahoo2013backtracking}. {\color{red}We use the fact that the survival probability density of a process described by stochastic matrix \(\mathbf{W}\) and starting in state \(i\) does not reach state \(j\) by time $t$ is given by~\cite{redner2001guide, sekimoto2021derivation}
\begin{equation}\label{survival}
\mathbb{S}(t, j\vert i) = \sum_{k\neq j} \bra{k} \exp( t \mathbf{W}) \ket{i} = 1- \bra{j} \exp( t \mathbf{W}) \ket{i},
\end{equation}
and rewrite this result for transitions rather than states.}

\begin{figure}[t!]
    \centering
    \includegraphics[width=.8\columnwidth]{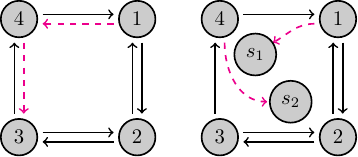}
    \caption{Left: example of a 4 states network with visible transitions \(\mathcal{L} = \{1\to 4, 4\to 3\}\) (dashed magenta). Right: network of the extended state space, where  visible transitions are redirected to auxiliary absorbing states $s_1$ and $s_2$.}
    \label{fig:sink}
\end{figure}

We consider a continuous-time Markov jump process  over an irreducible network  of discrete states \(\Omega = \{1,2,\ldots ,N\}\). We introduce auxiliary absorbing states \(s_i\) (sinks) for \(i\in [1, \abs{\mathcal{L}}]\) to account for the occurrence of every transition \(\ell \in \mathcal{L}\) separately. The \(\abs{\Omega_\text{ex}} \times \abs{\Omega_\text{ex}}\) stochastic matrix  \(\mathbf{W}_\text{ex} \) associated with the dynamics over the extended state space \(\Omega_\text{ex} \coloneqq \Omega \cup \{s_1, \ldots , s_{\abs{\mathcal{L}}}\}\) is such that {\color{red}every element of the visible set \(\ell \in \mathcal{L}\), a visible transition, is redirected to point towards its associated sink \(s_\ell\)} (cf. Fig.~\ref{fig:sink}) and since the sink is an absorbing state we set \([\mathbf{W}_\text{ex} ]_{j,s_i}=0\) for all $j$ in $\Omega$. Following our notation, the sources of visible transitions {\color{red}\(\lbra{\ell}\)} are preserved while the targets are redirected to the respective sinks {\color{red}\(\lket{\ell} \to \ket{s_\ell}\)}.

The extended matrix \(\mathbf{W}_\text{ex}\) has four blocks, the top-left block is the survival matrix \(\mathbf{S}\) with size \(\abs{\Omega} \times \abs{\Omega}\) and both blocks to the right are zero matrices. The bottom-left block \(\mathbf{L}\) has size \(\abs{\mathcal{L}} \times \abs{\Omega}\) and contains the redirected transitions, mathematically it is expressed as \(\mathbf{L} =\sum_{j=1}^{\abs{\mathcal{L}}} \lbra{\mathcal{L}_j} \mathbf{W}^\mathsf{T} \lket{\mathcal{L}_j} \ket{j} \lbra{\mathcal{L}_j}\), {\color{red}where the sum \(\sum_{j=1}^{\abs{\mathcal{L}}}\) runs through every element of the visible set of transitions}. For the example in Fig.~\ref{fig:sink} the extended stochastic matrix is

\begin{equation}
    \mathbf{W}_\text{ex} = \quad
        \begin{blockarray}{c c c c c c}
        {\color{gray}1} & {\color{gray}2} & {\color{gray}3} & {\color{gray}4} & {\color{gray}s_1} & {\color{gray}s_2} \\
        \begin{block}{(c c c c | c c)}\relax
        W_{11} & W_{12} & W_{13} & W_{14} & 0 & 0 \\\relax
        W_{21} & W_{22} & W_{23} & W_{24} & 0 & 0 \\\relax
        W_{31} & W_{32} & W_{33} & 0 & 0 & 0 \\
        0 & W_{42} & W_{43} & W_{44} & 0 & 0 \\
        \cmidrule{1-6}
        W_{41} & 0 & 0 & 0 & 0 & 0 \\
        0 & 0 & 0 & W_{34} & 0 & 0 \\
        \end{block}
        \end{blockarray}
\end{equation}

Since the last columns are zero, any power \(n\geq 1\) of the matrix has the property that its left blocks only depend on powers of themselves, and the right blocks remain zero
\begin{eqnarray}
    [ \mathbf{W}_\text{ex}^n]_{i,j} &=& \sum_{k\in\Omega} [ \mathbf{W}_\text{ex}]_{i,k}[ \mathbf{W}_\text{ex}^{n-1}]_{k,j} \nonumber\\
    &&+  \sum_{k'=1}^\abs{\mathcal{L}} {[ \mathbf{W}_\text{ex}]}_{i,\Omega + k'}[     \mathbf{W}_\text{ex}^{n-1}]_{\Omega + k',j} \nonumber\\
    &=& \sum_{k\in\Omega} [ \mathbf{W}_\text{ex}]_{i,k}[ \mathbf{W}_\text{ex}^{n-1}]_{k,j},
\end{eqnarray}
where the second equality follows from \([ \mathbf{W}_\text{ex}]_{i,\Omega + k'} =0\) for any \(i\) and \(1\leq k' \leq \abs{\mathcal{L}}\). Hence its matrix exponential \(\exp(\mathbf{W}_\text{ex})=\sum_{k=0}^\infty \mathbf{W}_\text{ex}^k/k!\) is such that its top-left block is its own exponential, i.e.
\begin{equation}\label{matrixexponential}
    \exp(t \mathbf{W}_\text{ex}) = \left( \begin{array}{c@{}|c@{}}
    \exp(t\mathbf{S}) \hspace{2pt} & 
    \mathbf{0}_{\abs{\Omega} \times \abs{\mathcal{L}}} \\
    \hline
    \mathbf{L} \mathbf{S}^{-1}(\exp(t\mathbf{S})-\mathbf{1}) \hspace{2pt} & \mathbf{1}_{\abs{\mathcal{L}}\times \abs{\mathcal{L}}} \\
\end{array} \right),
\end{equation}
where the last column has a non-square block of zeroes and one \(\abs{\mathcal{L}}\times \abs{\mathcal{L}}\) identity matrix.

We are interested in the case where a visible transition was performed at time zero, thus the initial state is \(\lket{\ell_i}\), \(\ell_i\in \mathcal{L}\), and the next visible transition \(\ell_{i+1} \in \mathcal{L}\) is performed by time \(t\), which is equivalent to the first-passage distribution. {\color{red}Hence the transition analogue of Eq.~\eqref{survival} is the survival density related to the respective sink \(s_{\ell_{i+1}}\):}
\begin{align}\label{survapp}
    \mathbb{S}(t, s_{\ell_{i+1}} \vert \ell_i ) &= 1- \bra{s_{\ell_{i+1}}} \exp( t \mathbf{W}_\text{ex}) \lket{\ell_i} \nonumber\\
    &= 1- \bra{s_{\ell_{i+1}}} \mathbf{L} \mathbf{S}^{-1}(\exp(t\mathbf{S})-\mathbf{1}) \lket{\ell_i},
\end{align}
where the last equality comes from the fact that the matrix element in question belongs to the bottom left block of Eq.~\eqref{matrixexponential}. The {\color{red}first-transition} distribution is given by the time derivative \( \mathbb{F} = -\partial_t \mathbb{S} \), hence
\begin{align}\label{fptapp}
    &\mathbb{F} (t, s_{\ell_{i+1}} \vert \ell_i ) \nonumber \\
    &=- \partial_t \mathbb{S} (t, s_{\ell_{i+1}} \vert \ell_i ) \nonumber\\
    &= \partial_t \bra{s_{\ell_{i+1}}} \mathbf{L} \mathbf{S}^{-1}(\exp(t\mathbf{S})-\mathbf{1}) \lket{\ell_i}\nonumber\\
    &= \bra{s_{\ell_{i+1}}} \mathbf{L} \exp(t\mathbf{S}) \lket{\ell_i} \nonumber\\
    &= \bra{s_{\ell_{i+1}}} \left(\sum_{j=1}^{\abs{\mathcal{L}}} \lbra{\mathcal{L}_j} \mathbf{W}^\mathsf{T} \lket{\mathcal{L}_j} \ket{j}\lbra{\mathcal{L}_j}\right) \exp(t\mathbf{S}) \lket{\ell_i}  \nonumber\\
    &= \lbra{\ell_{i+1}} \mathbf{W}^\mathsf{T} \lket{\ell_{i+1}} \lbra{\ell_{i+1}} \exp(t\mathbf{S}) \lket{\ell_i},
\end{align}
which provides the desired result
\begin{equation}
    P(t, \ell_{i+1} \vert \ell_i) = \lbra{\ell_{i+1}} \mathbf{W}^\mathsf{T} \lket{\ell_{i+1}} \lbra{\ell_{i+1}} \exp(t\mathbf{S}) \lket{\ell_i}
\end{equation}
for the joint probability density that a transition $\ell_{i+1}$ happens at a time $t$ given that $\ell_i$ was performed at time zero, and no other visible transition happened in between. $\blacksquare$

\section{Combinatorics of ring networks}\label{sec:appcombinatorics}

For a ring topology with reversible edges as described in Section \ref{sec:ring} we need to evaluate the coefficient $C_{\transR,\transR}^{\vec{k}}$. We observe that:

\begin{itemize}
    \item There are $\prod_{i=1}^{N-1} (k_i+1)$ possible backbones since there are $k_i+1$ clockwise edges between states $i$ and $i+1$. When the dynamics is occurring the chosen backbone is now a set of prohibited edges that are saved for last in order to make sure that the cycle will be completed.
    \item When the system is at state $1<i<N$ for the first time it can choose between $k_i$ clockwise and $k_{i-1}$ counterclockwise {\color{red}edges}. Next time $i$ is visited there will be one less way out of it since one transition was already used, and so on. Therefore we have the contribution $(k_i+k_{i-1})!$ from each state.
    \item The contribution of states 1 and $N$ are respectively $k_1!$ and $k_{N-1}!$ since {\color{red}from there it is only possible to jump in one direction.}
    \item Edges with the same direction and connecting the same pair of states are indistinguishable, so we have to divide everything by their number of permutations $(k_i+1)!k_{i}!$.
\end{itemize}

By gathering every contribution discussed above, we obtain
\begin{equation}
    C_{\transR,\transR}^{\vec{k}} = \prod_{i=2}^{N-1} \binom{k_i+k_{i-1}}{k_{i}}.
\end{equation}

For the alternated case notice that 
\begin{itemize}
    \item There are $\prod_{i=1}^{M} k_i$ possible backbones.
    \item When the system is at state $1<n<M+1$ for the first time it can choose between $k_i$ clockwise and $k_{i-1}-1$ counterclockwise edges (one is from the backbone). Next time $i$ is visited there will be one less way out of it since one transition was already used. Therefore we have the contribution $(k_i+k_{i-1}-1)!$ of each state.
    \item The contribution of states 1 and $M+1$ are respectively $k_1!$ and $(k_{M}-1)!$.
    \item Due to the indistinguishability of edges everything is divided by $k_n!^2$.
\end{itemize}
Thus leads to the coefficient
\begin{equation}
    C_{\transR,\transL}^{\vec{k},M} = \prod_{i=2}^{M} \binom{k_i+k_{i-1}-1}{k_i }.
\end{equation}

\section{Diagrammatic approach}\label{diagram}

The division \(\widehat{P}(s\vert \transR,\transR)/\widehat{P}(0\vert \transR,\transR)\) involves a product of \(\widehat{\pi}_{i-1,i}(s)/(\widehat{\pi}_{i-1,i}(0)/\), which by definition is \(W_{ii}/(W_{ii}+s)\). Therefore
\begin{equation}
    \frac{\widehat{P}(s\vert \transR,\transR)}{\widehat{P}(0\vert \transR,\transR)} = \left( \prod_{i=1}^N \frac{W_{ii}}{W_{ii}+s}\right) \frac{\biggr[\prod_{i=1}^{N-1} 1- \Theta [x_i] \biggr]_{s=0}}{\prod_{i=1}^{N-1} 1- \Theta [x_i]}
\end{equation}
and
\begin{equation}
    \frac{\widehat{P}(s\vert \transL,\transL)}{\widehat{P}(0\vert \transL,\transL)} = \left( \prod_{i=1}^N \frac{W_{ii}}{W_{ii}+s}\right) \frac{\biggr[\prod_{i=1}^{N-1} 1- \Xi (x_i) \biggr]_{s=0}}{\prod_{i=1}^{N-1} 1- \Xi (x_i)}.
\end{equation}
The first factors in the right-hand side of both equations above are the same, and using a diagrammatic approach we show that the second one is also the same.

Recall that {\color{red}the continued fraction generators used in Section~\ref{sec:ring} are defined as} \(\Theta[x_{i+1}] = x_{i+1}/(1-\Theta[x_{i}])\), \(\Theta[x_1] = x_1\), generates a continued fraction. To evaluate the monomial \(\prod_i (1-\Theta[x_i])\) we notice that if we pick two consecutive terms they simplify to
\begin{equation}\label{eq:paths}
    (1-\Theta[x_{i+1}])(1-\Theta[x_{i}])= 1- \Theta[x_{i}] - x_{i+1},
\end{equation}
which means that each multiplication of consecutive terms will lead to two terms: one that is only \((1-\Theta[x_i])\) and the other \(-x_{i+1}\). In other words there are two possible paths that will be added up to evaluate the whole product.

This branching procedure can be portrayed by the diagram below where each node \(j\) represents the value of \(\prod_{i=1}^j (1-\Theta[x_i])\). Notice that at each transition there are two arrows arriving, one representing the multiplication by \(+1\) and the other by \(-x_j\), as discussed.

\FloatBarrier
\begin{figure}[ht]
    \centering
    \includegraphics[width=.9\columnwidth]{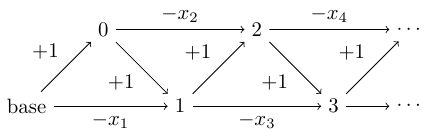}
\end{figure}
\FloatBarrier

The value of the monomial at each transition can be evaluated as the sum of all paths starting from the base and reaching such transition, weighted by the product of the weight of all edges involved.

To illustrate that consider the node 2:

\FloatBarrier
\begin{figure}[ht]
    \centering
    \includegraphics[width=.9\columnwidth]{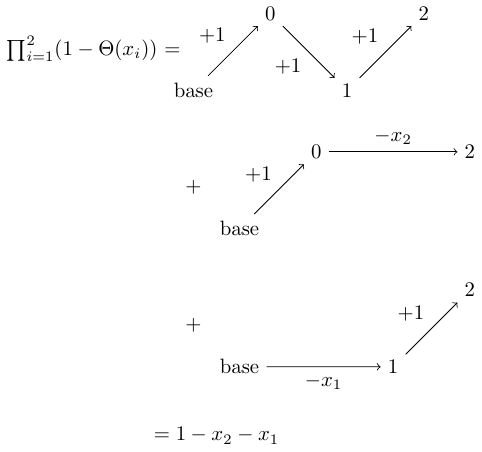}
\end{figure}
\FloatBarrier

For the reversed fraction generator \(\Xi[x_i]\) {\color{red}the analogue of Eq.~\eqref{eq:paths} is}
\begin{equation}
    (1-\Xi[x_{i+1}])(1-\Xi[x_{i}]) = 1-\Xi[x_{i+1}] - x_i,
\end{equation}
which means that the diagram has the same structure and, more importantly, it is covered in the backwards direction. The arrows point in the other direction but the weights remain unchanged, leading to the final result when \(\prod_i (1-\Theta[x_i])\) and \(\prod_i (1-\Xi[x_i])\) start and end at the same points, which implies
\begin{equation}
    \prod_{i=1}^{N-1} (1-\Theta[x_i]) = \prod_{i=1}^{N-1} (1-\Xi[x_i]).
\end{equation}
This property guarantees the Haldane-like equality Eq.~\eqref{eq:haldane}.

\section{Irreversibility in  transition time series}\label{infdetails}
 
We now evaluate analytically the rate of irreversibility in the time series $\Gamma^{\mathcal{L}}_t$ which is defined by Eq.~\eqref{inf_definition}, copied here for convenience:
\begin{equation}
    \sigma_{\mathcal{L}} = \lim_{\tau\to\infty} \frac{1}{\tau}D\left( P[ \Gamma_\tau^{\mathcal{L}} ] ||P[ \overline{\Gamma}_\tau^{\mathcal{L}} ]\right),
\end{equation}
where the probability of a trajectory can be expressed as the joint probability of all random variables involved, the visible transitions and inter-transition times
\begin{equation}
    P[ \Gamma_\tau^{\mathcal{L}} ] = P(t_0,\ell_0, t_1, \ell_1, \ldots) \equiv P(\vec{t}, \vec{\ell}).
\end{equation}

With no further assumptions, the path probability of the sequence of transitions can be cast as the product of the sequence probability and inter-transition times as \(P[\Gamma_\tau^{\mathcal{L}} ] = P(\vec{\ell}) P(\vec{t}\vert \vec{\ell})\). {\color{red}Since the underlying process is Markovian and there are no transitions with the same source and target, the sequence of transitions is also Markovian, \(P(\ell_i \vert \ell_{i-1}, \ldots, \ell_0) = P(\ell_i \vert \ell_{i-1})\), thus the probability of a sequence of transitions in a trajectory is}
\begin{equation}
    P(\vec{\ell}) = P(\ell_0) \prod_{i=1}^n P(\ell_i\vert \ell_{i-1}),
\end{equation}
where \(n\) is the total number of transitions within \(\Gamma_\tau^{\mathcal{L}}\). On the other hand, the probability of a sequence of inter-transition times conditioned to the occurrence of a given sequence of transitions reads
\begin{equation}
    P(\vec{t}\vert \vec{\ell}) = \prod_{i=1}^n P(t_i\vert \ell_{i-1}, \ell_{i}).
\end{equation}

Therefore the inferred entropy production rate reads
\begin{align}\label{firstandsecond}
    \sigma_{\mathcal{L}} &= \lim_{\tau\to\infty} \frac{1}{\tau} \sum_{\Gamma_\tau^\mathcal{L}} P[\Gamma_\tau^\mathcal{L}] \ln\frac{P[\Gamma_\tau^\mathcal{L}]}{P[\overline{\Gamma}_\tau]} \nonumber\\
    &= \lim_{\tau\to\infty} \frac{1}{\tau} \sum_{\vec{\ell}}\int\mathrm{d}t_0\cdots\mathrm{d}t_n \nonumber\\
    &\phantom{=}\biggr\lbrace P[\Gamma] \ln\frac{P[\vec{\ell}]}{P[\overline{\vec{\ell}}]}
    +P[\Gamma] \ln\frac{P[\vec{t}\vert \vec{\ell}]}{P[\overline{\vec{t}}\vert \overline{\vec{\ell}}]} \biggr\rbrace.
\end{align}
The first term in~\eqref{firstandsecond} is
{\color{red}
\begin{align}
    &\lim_{\tau\to\infty} \frac{1}{\tau} \sum_{\vec{\ell}}\int\mathrm{d}\vec{t} P[\Gamma_\tau^\mathcal{L}] \ln \frac{P(\ell_0)P(\ell_1\vert \ell_0)\cdots} {P(\overline{\ell}_n) P(\overline{\ell}_{n-1}\vert \overline{\ell}_n)\cdots} \nonumber\\
    &= \lim_{\tau\to\infty} \frac{1}{\tau} \sum_{\vec{\ell}} P[\vec{\ell}] \biggr\lbrace \ln P(\ell_0) +  \ln \frac{P(\ell_1 \vert \ell_0)} {P(\overline{\ell}_0\vert \overline{\ell}_1)} + \ldots \biggr\rbrace \nonumber\\
    &= \traff \sum_{\ell, \ell' \in \mathcal{L}} P(\ell \vert \ell') P(\ell') \ln \frac{P(\ell \vert \ell')} {P(\overline{\ell'}\vert \overline{\ell})} \nonumber\\
    &\eqqcolon \sigma_\ell.
\end{align}
The second term in~\eqref{firstandsecond} reads
\begin{align}
    &\lim_{\tau\to\infty} \frac{1}{\tau} \sum_{\vec{\ell}} \int\mathrm{d}\vec{t} P[\Gamma_\tau^\mathcal{L}] \ln\frac{P(t_1\vert \ell_0,\ell_1) \cdots} {P(t_n\vert \overline{\ell}_{n}, \overline{\ell}_{n-1})\cdots}\nonumber\\
    &=\lim_{\tau\to\infty} \frac{1}{\tau} \sum_{\vec{\ell}}P[\vec{\ell}] \biggr\lbrace \int\mathrm{d}t_1 P(t_1\vert \ell_0,\ell_1) \ln\frac{P(t_1\vert \ell_0, \ell_1)} {P(t_1 \vert \overline{\ell}_{1},\overline{\ell}_0 )}+\ldots\biggr\rbrace\nonumber\\
    &= \traff \sum_{\ell, \ell' \in \mathcal{L}} P(\ell \vert \ell') P(\ell')  D\left[ P(t\vert \ell', \ell) \vert \vert P(t\vert \overline{\ell}, \overline{\ell'}) \right] \nonumber\\
    &\eqqcolon \sigma_t.
\end{align}
}

For the special case of a single visible transition that  can only take values  \(\mathcal{L} = \{\transR,\transL\}\), {\color{red}the time reversed of \(\transR\) is \(\transL\) and vice-versa.} Inference of entropy production rate's first term simplifies to
\begin{equation}\label{firstterm}
    \sigma_\ell = \traff \biggr[ P(\transR \vert \transR) P(\transR)- P( \transL\vert \transL)P(\transL)\biggr] \ln\frac{P(\transR \vert \transR)}{P(\transL \vert \transL)}.
\end{equation}

The stationary occupation probability vector can be found by
\begin{equation}
    p_\infty(j) = (-1)^{i+j} \mathrm{det}(\mathbf{W}_{\backslash (i,j)}),\ \forall i
\end{equation}
since \(\mathrm{det}(\mathbf{W}) = \sum_i W_{ij} (-1)^{i+j} \mathrm{det}(\mathbf{W}_{\backslash (i,j)}) =0 \) and \(\sum_j W_{ij} p_\infty(j) = 0\) for every transition matrix defining a Markov chain. Lastly, by the construction of the survival matrix \(\mathbf{S}\), we observe that \(\mathbf{S}_{\backslash (i,i)} = \mathbf{W}_{\backslash (i,i)} \) and \(\mathbf{S}_{\backslash (j,j)} = \mathbf{W}_{\backslash (j,j)} \). Now we are in the position to show that
\begin{align}
    P(\transR\vert \transR) &= 1- P(\transL\vert \transR) \nonumber\\
    &= 1- \lbra{\transL}\mathbf{W}^\mathsf{T}\lket{\transL} \lbra{\transL} \mathbf{S}^{-1} \lket{\transR} \nonumber\\
    &= 1- \lbra{\transL}\mathbf{W}^\mathsf{T}\lket{\transL} \frac{\mathrm{det} (\mathbf{S}_{\backslash (2,2)}) }{\mathrm{det}(\mathbf{S})} \nonumber\\
    &= 1- \lbra{\transL}\mathbf{W}^\mathsf{T}\lket{\transL} \frac{\mathrm{det} (\mathbf{W}_{\backslash (2,2)}) }{\mathrm{det}(\mathbf{S})} \nonumber\\
    &= 1- \lbra{\transL}\mathbf{W}^\mathsf{T}\lket{\transL} \frac{ \lbra{\transL} p_\infty \rangle }{ \mathrm{det}(\mathbf{S})} \nonumber\\
    &=1- \frac{\traff P(\transL)}{\mathrm{det}(\mathbf{S})}
\end{align}
and analogously we obtain \(P(\transL\vert \transL )= 1- \traff P(\transR)/\mathrm{det}(\mathbf{S})\). Plugging this latter into Eq.~\eqref{firstterm} it simplifies to
\begin{equation}
    \traff [P(\transR) - P(\transL)] \ln \frac{P(\transR \vert \transR)}{P(\transL \vert \transL)} \eqqcolon J_\mathcal{L} A_\text{eff},
\end{equation}
which is Eq.~\eqref{epr_alpha} in the main text. The factor \(\traff [P(\transR) - P(\transL)]\) is the definition of the flux through the observed transition \(J_\mathcal{L}\), suggesting the definition of the second factor as the effective affinity \(A_\text{eff}\).

Following the same reasoning, the inter-transition times contribution simplifies to
\begin{align}\label{secondterm}
    \sigma_t =&\traff P(\transR\vert\transR)P(\transR) D[P(t\vert \transR, \transR) \vert \vert P(t\vert \transL, \transL)] \nonumber\\
    &+\traff P(\transL\vert\transL)P(\transL) D[P(t\vert \transL, \transL) \vert \vert P(t\vert \transR, \transR)],
\end{align}
which is Eq.~\eqref{epr_t} in the main text. The sum of Eq.~\eqref{firstterm} and \eqref{secondterm} results in the entropy production rate inferred by an observer who only accesses two opposite transitions between a single pair of states.

{\color{red}\section{Kullback-Leibler divergence from finite data}\label{app:convergence}

Estimation of   Kullback-Leibler divergences between distributions of continuous random variables, such as that present in \(\sigma_t\), from time series is not a straightforward task, as it can lead to systematic errors and statistical biases~\cite{PhysRevE.69.066138, Bonachela_2008, PhysRevE.85.031129}. 
Furthermore, inference schemes to deal with finite data have been largely explored in the analytical sense, as discussed in this paper, hence the need for accurate estimators.}

\begin{figure}[t!]
    \centering
    \includegraphics[width=0.95\columnwidth]{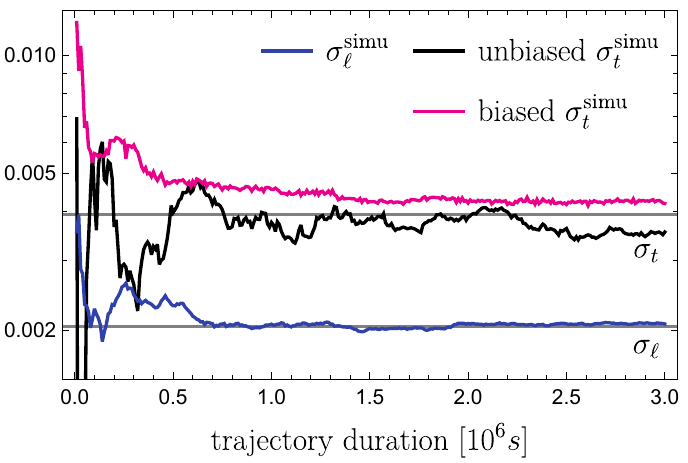}
    \caption{{\color{red}Convergence analysis of the two contributions to the entropy production rate inferred from numerical simulations \(\sigma_{\mathcal{L}}^\text{simu}\) for the model shown in Fig.~\ref{fig:grid2}] with bias parameter 8.6. The magenta line (top) is the inter-transition times' Kullback-Leibler divergence \(\sigma_t^\text{simu}\) evaluated by the biased method from histogram counting. The black line (middle) is the inter-transition times' Kullback-Leibler divergence \(\sigma_t^\text{simu}\) obtained using the unbiased method \cite{4595271, Harunari_KLD_estimation}. The blue line (bottom) is the estimate of the Kullback-Leibler divergence from transitions' occurrence statistics from the \(\sigma_\ell^\text{simu}\). The analytical values of both \(\sigma_t\) and \(\sigma_\ell\) are shown in horizontal gray lines.}}
    \label{fig:split}
\end{figure}

{\color{red}The most intuitive approach involves estimating the probability distributions (here $P$ and $Q$) via standard histogram counting methods of the data collected from an experiment or simulation, and later approximating the integral \(D[P(x)\vert \vert Q(x)] = \int {\rm d}x P(x) \ln P(x)/Q(x)\simeq \sum_i \mathsf{P}_i \ln (\mathsf{P}_i/\mathsf{Q}_i)\), with $\mathsf{P}_i$ and $\mathsf{Q}_i$ the probability for the data to fall in the $i-$th bin.  This approach however  leads to a biased estimate of the Kullback-Leibler divergence, as shown in previous work~\cite{PhysRevE.69.066138, Bonachela_2008, PhysRevE.85.031129}. A method developed in Ref.~\cite{4595271} explores an alternative, unbiased estimation method for  the Kullback-Leibler divergence bias-free which is based on the comparison between the cumulative distributions of two independent data sets generated by $P$ and $Q$. This method was adapted to the estimate of the inter-transition-time Kullback-Leibler divergences shown in the analysis of simulated results throughout the Main Text. We made our  code open-source and   available in Ref.~\cite{Harunari_KLD_estimation}, with further details and illustrations of generating visible transitions' time-series, and evaluating Kullback-Leibler divergences and \(\sigma_\mathcal{L}\).

Briefly, the method consists of taking two finite data sets sampled from two independent processes with distributions \(P(x)\) and \(Q(x)\). Linear interpolations \(F_c(x)\) and \(F_c(x)\) of their associated  empirical cumulative distributions are obtained for small enough \(\epsilon\) which are used in the following estimate that was shown to  converge to the Kullback-Leibler divergence Ref.~\cite{4595271}
\begin{equation}\label{eq:PerezCruzEstimator}
    \lim_{n\to \infty} \left[\frac{1}{n} \sum_{i=1}^n \ln \left(\frac{F_c(X_i) - F_c(X_i - \epsilon)}{F_c(X_i) - F_c(X_i - \epsilon)}\right) -1\right]  = D[P(x)\vert \vert Q(x)],
\end{equation}
where \(n\) is the number of points \(X_i\) from the data with distribution \(P(x)\).

Figure~\ref{fig:split} shows the converge of \(\sigma_\ell\) and \(\sigma_t\) estimates as the number of data points in the time series increases. The value of \(\sigma_\ell^\text{simu}\) has a fast convergence using the empirical frequencies of transitions. For $\sigma_t$  however the estimate depends strongly on the method used as inter-transition times are continuous random variables.  By estimating the probability distributions with kernel density estimations methods and numerically evaluating Kullback-Leibler divergence's integral, we obtain \(\sigma_t^\text{simu}\) (magenta line) which leads to a significant bias above the expected value. On the other hand, using Eq.~\eqref{eq:PerezCruzEstimator} through the resources in Ref.~\cite{Harunari_KLD_estimation}, the estimate \(\sigma_t^\text{simu}\) (black line) displays no evident sign of statistical bias above or below the analytical value of $\sigma_t$. This result motivated the usage of Eq.~\eqref{eq:PerezCruzEstimator} as our estimate for all the Kullback-Leibler divergences in this work.
}

\vfill

\bibliography{references}
\end{document}